\documentclass[11pt,a4paper,openright,twoside]{article}
\usepackage[left=1in,right=1in]{geometry}

\usepackage[english]{babel}

\usepackage{verbatim} %%per begin comment

\usepackage{amsmath}
\usepackage{amssymb}
\usepackage{amsfonts}
\usepackage{amsthm}
\usepackage{amscd}
\usepackage{latexsym}
\usepackage{amsthm}
\usepackage{mathtools}
\usepackage{braket}
\usepackage{enumerate}  % old package, new version is enumitem
\usepackage{cite}
\usepackage{booktabs}
\usepackage{array}
\usepackage{url}
\usepackage{paralist}
\usepackage{stmaryrd}
\usepackage{color}
\usepackage{longtable}
\usepackage{multicol}
\usepackage{csquotes}
 \usepackage{mathdots}
\usepackage{tikz}% for tikzpicture
\usepackage{tikz-cd}
\usepackage{pgfplots} % for tikz
\usepackage{algorithm}
\usepackage{algpseudocode}
\usepackage{float}    % for figure placement
\usepackage{hyperref} % for href
\usepackage{xcolor}
\usepackage{forest}
\usepackage{makecell}

\usepackage[thinlines,thiklines]{easybmat}

\numberwithin{equation}{section}

\newtheorem{theorem}{Theorem}[section]
\newtheorem*{theorem*}{Theorem}
\newtheorem{lemma}[theorem]{Lemma}
\newtheorem*{lemma*}{Lemma}
\newtheorem{corollary}[theorem]{Corollary}
\newtheorem{proposition}[theorem]{Proposition}
\newtheorem*{proposition*}{Proposition}
\theoremstyle{definition}
\newtheorem{definition}[theorem]{Definition}
\newtheorem*{definition*}{Definition}
\newtheorem{remark}[theorem]{Remark}
\newtheorem*{remark*}{Remark}
\newtheorem{example}[theorem]{Example}
\newtheorem*{example*}{Example}

\newtheorem*{exercise*}{Exercise}

\newtheorem*{algorithmm*}{Algorithm}

\newcommand{\kk}{\Bbbk} % Notazione per il campo
\newcommand{\NN}{\mathbb{N}}
\newcommand{\ZZ}{\mathbb{Z}}
\newcommand{\QQ}{\mathbb{Q}}

\newcommand{\TT}{\mathbb{T}}
\newcommand{\FF}{\mathbb{F}}

\newcommand{\mcF}{\mathcal{F}}

\newcommand{\GL}{\operatorname{GL}}
\newcommand{\Lex}{\operatorname{Lex}}
\newcommand{\DegLex}{\operatorname{DegLex}}
\newcommand{\DL}{\operatorname{DL}}
\newcommand{\DegRevLex}{\operatorname{DegRevLex}}
\newcommand{\DRL}{\operatorname{DRL}}
\newcommand{\LT}{\operatorname{LT}}

\newcommand{\LM}{\operatorname{LM}}
\newcommand{\LC}{\operatorname{LC}}
\newcommand{\Supp}{\operatorname{Supp}}
\newcommand{\NF}{\operatorname{NF}}

\newcommand{\lcm}{\operatorname{lcm}}
\newcommand{\Pair}{\operatorname{Pair}}
\newcommand{\Pairs}{\operatorname{Pairs}}

\newcommand{\Sel}{\operatorname{Sel}}
\newcommand{\List}{\operatorname{List}}
\newcommand{\Done}{\operatorname{Done}}

\newcommand{\multiVarRed}{\operatorname{\textcolor[HTML]{006EB8}{multiVarRed}}}
\newcommand{\multiVarDiv}{\operatorname{\textcolor[HTML]{006EB8}{multiVarDiv}}}
\newcommand{\symPre}{\operatorname{\textcolor[HTML]{006EB8}{symPre}}}
\newcommand{\sPol}{\operatorname{\textcolor[HTML]{006EB8}{sPol}}}
\newcommand{\Red}{\operatorname{\textcolor[HTML]{006EB8}{Red}}}
\newcommand{\Ffour}{\operatorname{\textcolor[HTML]{006EB8}{F4}}}
\newcommand{\shalf}{\operatorname{\textcolor[HTML]{006EB8}{ s\_half}}}
\newcommand{\rowsp}{\operatorname{rowsp}}
\newcommand{\asmatrix}{\operatorname{matrix}}
\newcommand{\rows}{\operatorname{rows}}
\newcommand{\interred}{\operatorname{\textcolor[HTML]{006EB8}{interred}}}
\newcommand{\AscOrdLT}{\operatorname{\textcolor[HTML]{006EB8}{AscOrdLT}}}

\def\0{{\bf 0}}

\algrenewcommand\algorithmicrequire{\textbf{Input:}}
\algrenewcommand\algorithmicensure{\textbf{Output:}}

\newcommand{\mycode}[1]{\textcolor{blue}{\texttt{#1}}}

\title{{\Huge Solving Polynomial Systems \\with Gr\"obner bases:\vspace{2mm} \\ An Introduction to F4 and FGLM     }}
\author{{\Large Anna Maria Bigatti}\vspace{1mm}\\ 
\vspace{1mm}{\Large Alessio Caminata}
\\ \vspace{1mm} {\Large Tor Kristian Ellingsen}
\\ \vspace{1mm}{\Large Evelina Lanteri}
\\ \vspace{1mm}{\Large Andrea Sanguineti}
\\ \vspace{1mm}{\Large Irene Villa}}
\date{ }
\begin{document}
	\maketitle

\newpage
%%%%%%%%%%%%%%%%%%%%%%%%%%%%%%%%%%%%%%%%%%%%%%%%%%%%%%%%%%%%%%%%%%%
%%%%%%%%%%%%%%%%%%%%%%%%%  Indice %%%%%%%%%%%%%%%%%%%%%%
\tableofcontents

\newpage
%%%%%%%%%%%%%%%%%%%%%%%%%%%%%%%%%%%%%%%%%%%%%%%%%%%%%%%%%%%%%%%%%%%
%%%%%%%%%%%%%%%%%%%%%%%%% Nuovo Capitolo %%%%%%%%%%%%%%%%%%%%%%
\section*{Notations}\label{chapternotation}
%\addcontentsline{toc}{section}{Notation}

\par We will adopt the following notations.

\begin{itemize}	
	\item With $\NN$ we denote the set of natural numbers including $0$, that is $\NN:=\{0,1,2,3,\dots\}$.
	\item  With $\NN^*$ we denote the set of natural numbers excluding $0$, that is $\NN^*:=\{1,2,3,\dots\}$.
	\item With $\ZZ$ we denote the set of integer numbers, that is $\ZZ:=\{\dots,-3,-2,-1,0,1,2,3,\dots\}$.
 \item With $\QQ$ we denote the set of rational numbers.
	\item With $\FF_q$ we denote the finite field with $q$ elements, $\FF_q^*=\FF_q\setminus\{0\}$.
\end{itemize}

In general, we will denote a field by $\kk$ and a polynomial ring in $n$ variables over $\kk$ by $\kk[x_1,\dots,x_n]$. We use lowercase letters to denote polynomials.\\
\begin{itemize}
\item With $|\alpha|$, for $\alpha=(\alpha_1,\ldots,\alpha_n)\in\NN^n$, we indicate the sum $\alpha_1+\cdots+\alpha_n$.
\item With $f\mid g$, for $f,g\in\kk[x_1,\ldots,x_n]$, we indicate that $f$ divides $g$.
  \item  With $ax^\alpha$ we denote the \textbf{monomial} $ax_1^{\alpha_1}\cdots x_n^{\alpha_n}$ where $\alpha = (\alpha_1 ,\ldots, \alpha_n) \in \mathbb{N}^n$ and $a \in \kk$. If $a = 1$ we call it \textbf{term}. 
  \item With $\TT\left(x_1, \dots, x_n\right)$ we denote the set of all terms in $\kk[x_1,\dots,x_n]$ that is  $ \TT\left(x_1, \dots, x_n\right)\ = \{ x^\alpha \mid \alpha \in \NN^n \}$. If there is no ambiguity, we will denote it simply by $\TT$.
  \item With $\Supp(f)$ we denote the set of terms that appears with non zero coefficient in the polynomial $f$. More precisely, if $f = a_1x^{\alpha_1} + \dots + a_m x^{\alpha_m}$ with $a_1a_2 \cdots a_m \neq 0$ then
  $$\Supp(f) = \left\{x^{\alpha_1}, \dots, x^{\alpha_m}\right\} .$$
  \item If $F \subset \kk[x_1,\dots,x_n]$ is a subset of polynomials, we denote by $\Supp(F)$ the set
  $$\Supp(F) = \bigcup\limits_{f \in F} \Supp(f),$$
  that consists of the union of all the terms in the support of all the polynomials in $F$.
  \item With $\deg(f)$ we denote the degree of the polynomial $f$, that is the maximum degree of the monomial in $\Supp(f)$, where $\deg(x^\alpha)=|\alpha|$. 
  \item The set of $m \times n$ matrices with coefficients in $\kk$ is denoted with $\mbox{M}_{m,n}(\kk)$. The set of $n \times n$ matrices with coefficients in $\kk$ will be denoted with $\mbox{M}_n(\kk)$. The set of $n \times n$ invertible matrices will be denoted with $\GL_n(\kk).$
  \item If $f \in \kk[x_1,\dots,x_n]$, with $\NF_<(f,I)$ we denote the multivariate complete division $$\multiVarDiv(f, G, <)$$ for $G$ a Gr\"obner basis of $I \subseteq \kk[x_1,\dots,x_n]$ w.r.t. the selected term order $<$, as explained in Section~\ref{sec:polynomialdivision}.

\end{itemize}

\vspace{5mm}

%%%%%%%%%%%%%%%%%%%%%%%%%%%%%%%%%%%%%%%%%%%%%%%%%%%%%%%%%%%%%%%%%%%
%%%%%%%%%%%%%%%%%%%%%%%%%  Sezione %%%%%%%%%%%%%%%%%%%%%%	
\newpage
\section*{Introduction}\label{chapterintroduction}
\addcontentsline{toc}{section}{Introduction}

\textit{These notes originate from a reading course held by the authors in the spring of 2024 at the Università di Genova.}
\vspace{3mm}

\par Suppose we have multivariate polynomials $f_1,\ldots,f_c$ with coefficients in a field $\kk$. We want to find the solutions (possibly in some extension field) of the corresponding polynomial system:

\begin{equation}{\label{sistema}}
	\begin{cases}
	$$f_1(x_1, \dots, x_n) = 0 \\
	\hspace*{1cm}\vdots \\
	f_c(x_1, \dots, x_n) = 0 $$
	\end{cases}
\end{equation}
This problem is sometimes referred to in the literature as the PoSSo (Polynomial System Solving) Problem and has significant applications in various areas of Mathematics. In particular, we are motivated by applications in Cryptography, where the security of several schemes can be reduced to a PoSSo problem over a finite field. In this context, numerical methods are not applicable, and one must rely on symbolic methods, such as Gr\"obner bases, introduced by Bruno Buchberger in his Ph.D. thesis \cite{Buch70, Buch06}.

In a nutshell, a Gr\"obner basis is a \emph{special} set of generators of an ideal in the polynomial ring $\kk[x_1,\dots,x_n]$ which depends both on the ideal itself and on a term order, i.e., a total order on the set of monic monomials of the polynomial ring. The key connection with the PoSSo Problem is provided by the lexicographic term order. Specifically, under suitable assumptions, the solution of a polynomial system as in \eqref{sistema} can be directly obtained from a lexicographic Gr\"obner basis of the ideal $(f_1,\dots,f_c)$.
Thus, solving a polynomial system can be reduced to computing a lexicographic Gr\"obner basis. However, it is often more efficient to divide this computation into two steps:
\begin{enumerate}
    \item Compute a Gr\"obner basis with respect to a degree-compatible term order.
    \item Convert this basis into a lexicographic Gr\"obner basis. 
\end{enumerate}
The first step can be performed using any algorithm for Gr\"obner basis computation, such as Buchberger’s original algorithm. However, a significant improvement in this process can be traced back to Lazard \cite{Laz83}, who introduced the idea of converting Gr\"obner basis computations into several instances of linear algebra problems using Macaulay matrices. This approach led to the development of new \emph{linear algebra based} algorithms, such as F4 \cite{F4paper} and F5 \cite{Fau02}, which have proven highly successful in many practical instances.
The second step can be performed using the Gr\"obner walk strategy \cite{Grobnerwalk} or, in the case of a zero--dimensional ideal (i.e., an ideal with only a finite number of solutions over $\overline{\kk}$) with the FGLM Algorithm.

The primary goal of these notes is to provide an introduction to the FGLM and F4 algorithms, which are presented in Section~\ref{sec FGLM} and Section~\ref{sec F4}, respectively.
Over time, these algorithms have been refined, modified and implemented in several computer algebra systems, such as CoCoA \cite{CoCoALib, CoCoA}, Macaulay2 \cite{M2}, Magma \cite{magma}, msolve \cite{msolve}, and Sage \cite{sage}. However, these implementations are often optimized for performance and, in some cases, are not even openly accessible. As a result, they may not be easy for newcomers to understand.
For this reason, we have chosen to present the basic \enquote{textbook} versions of both FGLM and F4, following the original ideas from \cite{FGLM} and \cite{F4paper}, respectively. For FGLM, we present an implementation in CoCoALib \cite{CoCoALib}, while, in the case of F4, we provide a basic implementation using Python and the Sage programming language (Section~\ref{sec:implementation}). These implementations closely follow the structure of the algorithms as presented in these notes. We hope they will be useful for readers who wish to explore the algorithms hands-on and experiment with them to gain a deeper understanding. 

\subsection*{Acknowledgements}
The authors were supported by the Italian PRIN2022 grant 2022J4HRR \enquote{Mathematical Primitives for Post Quantum Digital Signatures}, by the MUR Excellence Department Project awarded to Dipartimento di Matematica, Università di Genova, CUPD33C23001110001, and by the European Union within the program NextGenerationEU.
T.K.~Elligsen was supported by the Erasmus+ Programme of the European Union and by the University of Bergen.

%%%%%%%%%%%%%%%%%%%%%%%%%%%%%%%%%%%%%%%%%%%%%%%%%%%%%%%%%%%%%%%%%%%
%%%%%%%%%%%%%%%%%%%%%%%%%  Sezione %%%%%%%%%%%%%%%%%%%%%%	

\newpage

\section{Preliminaries on Gr\"{o}bner Bases}

In this section, we present definitions and some preliminary results on Gr\"obner bases and their connection to polynomial system solving.
Our goal is not to provide an exhaustive introduction but rather to offer a concise exposition that establishes the notation and arguments for what follows. We include only selected proofs and refer the reader to comprehensive textbooks or notes such as \cite{CLS97, Faug-notes, KR1, KR2} for a more in-depth introduction.

\subsection{Term Ordering and Gr\"obner Bases}
We fix a field $\kk$ and a polynomial ring $R=\kk[x_1,\ldots,x_n]$. 
If $f$ is a polynomial in one variable, there is a standard way to order the terms in $\Supp(f)$, namely by their degree, which provides a canonical way to write $f$. This ordering is also useful for comparing polynomials, even when they contain a large number of terms. However, when $n > 1$, the situation becomes more complex: there is no immediate or natural way to order the terms of $R$. So, what do we mean by \textit{ordering} in the case of the terms of a polynomial ring $R$?
\begin{definition}\label{term_order_def}
A {\bf term order} $\tau$ on $R$ is a total relation order $\leq_\tau$ on $\TT$ such that 
\begin{enumerate}
    \item For all $m_1,m_2,m \in \TT \text{ if }m_1 \le_{\tau} m_2$ then $m_1m \le_{\tau} m_2m$.
    \item For all $m_1,m_2 \in \TT \text{ such that } m_1 \mid m_2$ then $m_1 \le_{\tau} m_2$.
\end{enumerate}
\end{definition}
\begin{remark}
If condition $1.$ of the definition holds, then the second condition is equivalent to:
\begin{enumerate}
    \item[$2'.$] For all $i \in \left\{1, \dots, n\right\}$, $1 <_\tau x_i$.
\end{enumerate}
\end{remark}

\begin{example}
	We present three important examples of term orders.
	\begin{itemize}		
		\item Lexicographic order ($\Lex$)\\
		$x^\alpha \geq_{\Lex} x^\beta$ if and only if the first non-zero entry from left to right of $\alpha-\beta \in \mathbb{Z}^n$ is positive.
		
		\item Degree Lexicographic ($\DegLex$, for short $\DL$)\\
		$x^\alpha \geq_{\DL} x^\beta$ if and only if $|\alpha| > |\beta|$ or $|\alpha| = |\beta|$ and $x^\alpha \geq_{\Lex} x^\beta$.
		
		\item Degree Reverse Lexicographic ($\DegRevLex$, for short $\DRL$)\\
		$x^\alpha \geq_{\DRL} x^\beta$ if and only if $|\alpha| > |\beta|$ or $|\alpha| = |\beta|$ and the first non-zero entry from right to left of $\alpha-\beta \in \mathbb{Z}^n$ is negative.		
	\end{itemize}
\end{example}

\begin{definition}
    We say that a term order $\tau$ is {\bf degree compatible} if for all $m_1,m_2 \in \TT$ such that $\deg(m_1)<\deg(m_2)$ then $m_1 <_{\tau} m_2$.
\end{definition}

\noindent We observe that $\DegLex$ and $\DegRevLex$ are degree compatible while $\Lex$ is not.

\begin{remark}
Term orders on $R$, and expecially those who are not degree-compatible, exhibit very unorthodox behaviors. For example, if we take $R = \kk[x,y,z]$ equipped with $\geq_{\Lex}$ term order ($x \geq_{\Lex} y \geq_{\Lex} z$), we get that $xy >_{\Lex} z^{1000}$. In fact, there exist infinitely many terms that are strictly smaller than $xy$:
$$xy >_{\Lex} z^n \ \mbox{for all} \ n \in \mathbb{N}^*.$$ 
\end{remark}

However, the following lemma tells us that, with every term order we choose, we cannot have infinite (strictly) descending sequences of terms. It is an easy consequence of the Noetherian property of the polynomial ring $R$.

\begin{lemma}\label{descending chains}
Let $\tau$ be a term order on $R$, then there are no infinite descending chains of terms with respect to $\tau$.
\end{lemma}
\begin{proof}
Suppose by contradiction that there is an infinite descending chain of terms
$$t_1 >_{\tau} t_2 >_{\tau} \cdots >_{\tau} t_i >_{\tau} \cdots$$
with respect to a fixed term order $\tau$ on $\mathbb{T}$. Consider the ideal 
$$I = (t_i \ | \ i \in \mathbb{N}^*)$$
generated by all the terms $t_i$. Since $R$ is Noetherian, there exists a finite set $J \subseteq \{t_i\}_{i\in\mathbb{N}^*}$ such that $I = (J)$. Let 
$$N := \max\{\ell \in \mathbb{N}^* \ | \ t_{\ell} \in J\}.$$
Now, let $j > N$; then there exists $t \in J$ such that $t | t_j$. So, by Definition \ref{term_order_def} and by construction, we have
$$t_j \geq_{\tau} t \geq_{\tau} t_N >_{\tau} t_j,$$
which provides the contradiction $t_j >_{\tau} t_j$. $\lightning$
\end{proof}

A term order $\tau$ gives us a way to write a polynomial $f \in R$ in a unique ordered way, that is
$$f = \displaystyle\sum_{i=1}^dc_it_i \text{ with } c_i\in\kk,\ t_i\in\TT,\ \text{and}\ t_1 >_\tau t_2 >_\tau \dots >_\tau t_d .$$
We will call $t_1$ the leading term of $f$, $c_1$ the leading coefficient of $f$ and $c_1t_1$ the leading monomial of $f$. More precisely, we have the following definition.
\begin{definition}
    Given a polynomial $f \in R$ and a term order $\tau$ on $R$, we define:
    \begin{itemize}[-]
    \item the \textbf{leading term} of $f$ w.r.t.\ the term order $\tau$ as $$\LT_{\tau}(f)= \displaystyle\max_{\tau} \Supp(f) .$$
    \item the \textbf{leading coefficient} $\LC_{\tau}(f)$ of $f$ w.r.t.\ the term order $\tau$ as the coefficient of the term $\LT_{\tau}(f)$ in $f$.
    \item the \textbf{leading monomial} of $f$ w.r.t.\ the term order $\tau$ as $$\LM_{\tau}(f)= \LC_\tau(f)\LT_\tau(f) .$$
    \end{itemize}
\end{definition}

\noindent Given a finite set of polynomials $F = \left\{f_1, \dots, f_r\right\}$, we denote:
$$
\LT_\tau(F) := \left\{\LT_\tau(f_1), \dots, \LT_\tau(f_r)\right\} .
$$
\begin{definition}
    Let $I \subseteq R$ be an ideal; we define the \textbf{leading term ideal} or \textbf{initial ideal} of $I$ as
    $$
        \LT_{\tau}(I):=(\LT_{\tau}(f) \mid f \in I\setminus\left\{0\right\}).
    $$
\end{definition}

\begin{remark}
Please note that if $F \subseteq R$ is a set or a list of polynomials, $\LT_{\tau}(F)$ is simply the set of all the leading terms of elements of $F$, while, if $I \subseteq R$ is an ideal, $\LT_{\tau}(I)$ denotes the ideal generated by the leading terms of all the non-zero elements of $I$.
\end{remark}

\begin{remark}
If $I = (f_1,\ldots,f_r)$,  in general, we have a strict inclusion $$(\LT_\tau(f_1),\ldots,\LT_\tau(f_r)) \subsetneq \LT_\tau(I).$$ 
\end{remark}

\begin{example}
    Let us consider $R= \mathbb{Q}[x,y]$, $I = (f_1,f_2)=(x^2+y^2,xy)$ and $\tau = \Lex$.     
    Then we have
    $(\LT(f_1),\LT(f_2)) = (x^2,xy)$. On the other hand, it is clear that $y^3 \notin (\LT(f_1),\LT(f_2))$, while
    $y^3 = y(x^2+y^2)-x(xy) \in I$ so $y^3 \in \LT(I)$.
\end{example}
When equality holds, then we say that $\{f_1,\ldots,f_r\}$ is a Gr\"{o}bner basis of $I$.
\begin{definition}\label{GBdef}
Let $I \subseteq R$ be an ideal, we say that the finite set $G = \left\{g_1, \dots, g_r\right\} \subseteq I$ is a \textbf{Gr\"obner basis} of $I$ w.r.t.\ the term order $\tau$ if
$$
    (\LT_\tau(g_1),\ldots,\LT_\tau(g_r)) = \LT_\tau(I).
$$
Furthermore $G$ is said to be {\bf reduced} if the following conditions hold:
\begin{enumerate}[(i)]
\item $\left\{\LT_\tau(g_1), \dots, \LT_\tau(g_r)\right\}$ minimally generates $\LT_\tau(I)$ .
\item For all $i \in \left\{1, \dots, r\right\}$, $\LC_\tau(g_i) = 1.$
\item For all $i \in \left\{1, \dots, r\right\}$, $\Supp(g_i) \cap \LT_\tau(I) = \left\{\LT_\tau(g_i)\right\}$
\end{enumerate}
\end{definition}
A Gr\"obner basis is not only a subset of $I$; it is not hard to see that it is a set of generators for the ideal.
\begin{lemma}
If $\left\{g_1, \dots, g_r\right\}$ is a Gr\"obner basis of the ideal $I$ (w.r.t.\ the term order $\tau$), then
$$
I = \left(g_1, \dots, g_r\right) .
$$
\end{lemma}
\begin{proof}
One inclusion is trivial, so we show the other. \\ Let $f \in I\setminus\{0\}$. Since $(\LT_\tau(g_1),\ldots,\LT_\tau(g_r)) = \LT_\tau(I)$, there exists $i_1 \in \{1,\dots,r\}$ such that $\LT_{\tau}(g_{i_1})\mid\LT_{\tau}(f)$. Let $f_1 := f$ and let $f_2 := f_1 - \frac{LC_{\tau}(f_1)}{LC_{\tau}(g_{i_1})} \frac{\LT_{\tau}(f_1)}{\LT_{\tau}(g_{i_1})}g_{i_1}$. Then either $f_2 = 0$ or $\LT_{\tau}(f_1) >_{\tau} \LT_{\tau}(f_2)$. If we are in the first case we have proved that $f \in (g_1,\dots,g_r)$. So, assume we are in the second case. Then, as before, there exists $i_2 \in \{1,\dots,r\}$ such that $\LT_{\tau}(g_{i_1})\mid\LT_{\tau}(f_2).$ So let $f_3 := f_2 - \frac{LC_{\tau}(f_2)}{LC_{\tau}(g_{i_2})} \frac{\LT_{\tau}(f_2)}{\LT_{\tau}(g_{i_2})}g_{i_2}$. Then either $f_3 = 0$ or $\LT_{\tau}(f_1) >_{\tau} \LT_{\tau}(f_2) >_{\tau} \LT_{\tau}(f_3)$. If we are in the first case we have proved that $f \in (g_1,\dots,g_r)$. Otherwise we repeat the process. Now, we observe that we cannot always fall in the second case, otherwise we would be able to construct an infinite descending sequence of terms
$$\LT_{\tau}(f_1) >_{\tau} \LT_{\tau}(f_2) >_{\tau} \cdots >_{\tau} \LT_{\tau}(f_i) >_{\tau} \cdots,$$
and this would contradict Lemma \ref{descending chains}. So, in the end we are able to write $f \in I$ as a combination of the elements $g_1,\dots,g_r$ with coefficients in $R$. This proves that $I \subseteq (g_1,\dots,g_r)$.
\end{proof}

\subsection{Polynomial Division and Polynomial Reduction}\label{sec:polynomialdivision}
One of the most important results for univariate polynomials is the existence of Euclidean division. In the multivariate case, however, the situation is more complex.
We now introduce a definition of polynomial division for multivariate polynomials.
\begin{definition}\label{division_with_remainder_def}
Let $\tau$ be a term order, $f, \, g_1, \dots, g_c \in R$, with $g_i \neq 0$ for $i=1, \dots, c$. A \textbf{division with remainder} of $f$ w.r.t.\ $g_1, \dots, g_c$ is an expression of the form
$$
f = q_1g_1+ \cdots + q_cg_c + r
$$ where $q_1,\ldots,q_c,r\in R$ are 
such that:
\begin{enumerate}[(1)]
    \item $q_i = 0$ or $\LT_\tau(q_ig_i) \leq \LT_\tau(f)$
    \item $r=0$ or $\LT_\tau(r) \notin \left( \LT_\tau(g_1), \dots, \LT_\tau(g_c) \right)$. 
\end{enumerate}
Furthermore, if $\Supp(r)\cap\left(\LT_\tau(g_1), \dots, \LT_\tau(g_c)\right) = \emptyset$, the division is said to be \textbf{complete}.
\end{definition}
It is important to notice that there is not a unique way to perform a division.
We now give an algorithm to \textit{reduce} a polynomial $f$ via the polynomials in $G = \left\{g_1, \dots, g_c\right\}$, that is to find the remainder $r$ of a division of $f$ w.r.t.\ $G$.

\begin{algorithm}[H]
		\caption{$\multiVarRed$}
            \label{PolynomialReduction}
		\begin{algorithmic}
			\Require $f \in R$ a polynomial, $G = [g_1,\dots,g_c]$ a list of polynomials, $<$ a term order.
			\Ensure a reduced polynomial			
			\While{$f \not= 0$ \textbf{and} $\exists$ $h \in F$ such that $\LT(h) \mid \LT(f)$}
            \State $k := \min\{i\in \{1,\dots,m\} \ \mbox{such that} \ \LT(f_i) \mid \LT(f)\}$
           \State $f := f - \frac{\LM(f)}{\LM(f_k)}f_k$; 
            \EndWhile             
            \State\Return $f$
		\end{algorithmic}
\end{algorithm}
The termination of Algorithm~\ref{PolynomialReduction} is ensured by the fact that there is no infinite descending chain of terms with respect to a term order (Lemma \ref{descending chains}). Moreover, if we set $r := \multiVarRed(f,G,<)$, then it is easy to see that
$$r - f \in (G).$$
Now, we have also that, if $r \not= 0$, then
$$\LT(r) \notin (\LT(G)),$$
so this is indeed the remainder of a division of $f$ w.r.t.\ the list $G$.
We note that it might still be the case that there exists $t \in \Supp(f) \setminus \LT(G)$ such that $\LT(f_i) \mid t$ for some $f_i \in G$, so this division is not complete in general. Another issue is that the $\multiVarRed$ strongly depends on the order of the polynomials in the list $G$.
To overcome the first problem, we introduce the total (or complete) reduction of a polynomial.
\begin{algorithm}[H]
		\caption{$\multiVarDiv$}
            \label{PolynomialTotalReduction}
		\begin{algorithmic}
			\Require $f \in R$ a polynomial, $G = [g_1,\dots,g_c]$ a list of polynomials, $<$ a term order.
			\Ensure a totally (or completely) reduced polynomial.
                    \State $p:=f, \quad p_0 := 0$                
                    \While{$p \not= 0$}
                    \State $p := \multiVarRed(p,G,<)$
                    \State $p_0 := p_0 + \LM(p)$
                    \State $p := p - \LM(p)$ 
                    \EndWhile             
                    \State\Return $p_0$
		\end{algorithmic}
\end{algorithm}
For the same reason as $\multiVarRed$, this algorithm terminates. \\
As for $\multiVarRed$, we have that
$r - f \in (G)$, moreover, now, if $$r := \multiVarDiv(f,G,<),$$ then we also have
$$\Supp(r) \cap (\LT(G)) = \emptyset.$$
This means that $r$ is the remainder of a complete division of $f$ w.r.t.\ $G$.

\begin{theorem}\label{thm:uniquedivision}
Let $I$ be an ideal and $<$ term order. Then, for every $f \in R$ there exist unique $g$ and $r$ such that $f=g+r$, where $g \in I$
 and either $r=0$
 or $\Supp(r) \cap (\LT_<(I)) = \emptyset$.
\end{theorem}
\begin{proof}
We consider a Gr\"obner basis $G$ of $I$ w.r.t.\ $<$. Applying the Algorithm \ref{PolynomialTotalReduction} to $f, G$ and $<$ gives us the existence of $g$ and $r$.
Suppose now that we can write $f = g_1 + r_1$ and $f = g_2 + r_2$ with $g_1, g_2 \in I$ and either $r_i=0$
 or $\Supp(r_i) \cap (\LT_<(I)) = \emptyset$ for $i = 1,2$.
 Then $r_1 - r_2 = (f-g_1)-(f-g_2) \in I$ but none of its monomials are in $\LT_<(I)$, since $\LT_<(I)$ is a monomial ideal this means that $r_1-r_2 = 0$ and the uniqueness follows.
\end{proof}
\noindent A consequence of Theorem~\ref{thm:uniquedivision} is that, when $G$ is a Gr\"obner basis of $I$, then the output of
$\multiVarDiv(f,G,<)$
is unique no matter the order of the polynomials in $G$. Moreover, in this case,
$$\multiVarDiv(f,G,<) = 0 \, \text{ if and only if } \, f \in (G).$$
From now on we will use the standard notation $\NF_<(f,I)$ to denote $$\multiVarDiv(f, G, <)$$ for $G$ a Gr\"obner basis of $I$ w.r.t.\ $<$. 

We have talked about the polynomial reduction and polynomial division. A generalization of these concepts is the one of \textit{interreduction}.

\begin{definition}\label{def:interreduction}
Let $F = [f_1,\dots,f_s] \subseteq R$ be a list of non-zero polynomials. We say that $F$ is \textbf{interreduced} w.r.t. the term order $\tau$ if for every $i =1,\dots, s$ we have
$$ \Supp(f_i) \cap (\LT(F \smallsetminus\{f_i\})) = \emptyset,$$
that is, no term in the support of any $f_i$ is reducible by the leading term of an $f_j, j \not= i$.
\end{definition}

Notice that, if $G = [g_1,\dots,g_r] \subseteq R$ is a Gr\"obner basis for $(G)$ such that $g_1,\dots,g_r$ are monic, pairwise different and interreduced, then $G$ is a reduced Gr\"obner basis (in the sense of Definition \ref{GBdef}).
 
If $F = [f_1,\dots,f_s]$ is not interreduced, there is a simple procedure called interreduction ($\interred$, Algorithm~\ref{interredltfirst}) to interreduce it, based on the $\multiVarDiv$ algorithm. The output of $\interred$ is a new list $F' = [f_1',\dots,f_{s'}']$ such that $(F) = (F')$, $(\LT(f_1),\dots,\LT(f_s)) \subseteq (\LT(f_1'),\dots,\LT(f_{s'}'))$ and all the polynomials in $F'$ are monic. To present the pseudo-code of the $\interred$ algorithm, we need a simple auxiliary function, called $\AscOrdLT$, that, given $H = [h_1,\dots,h_c] \subseteq R$ and $<_{\tau}$ a term order, returns the list $[h_{\sigma(1)},\dots,h_{\sigma(c)}] \subseteq R$ such that $\sigma \in {S}_c$ is a permutation that yields
$$\LT(h_{\sigma(1)}) \leq_{\tau} \cdots \leq_{\tau} \LT(h_{\sigma(c)}).$$

\begin{algorithm}%[H]
		\caption{$\interred$}
            \label{interredltfirst}
		\begin{algorithmic}
			\Require $F = [f_1,\dots,f_s] \subseteq R$ a list of non-zero pair-wise distinct polynomials,\\ \ \quad \quad  $<_{\tau}$ a term order.
			\Ensure an interreduced list of polynomials $F' = [f_1',\dots,f_{s'}'] \subseteq R$ such that \\ \ \quad \quad \quad $(F) = (F')$ and $(\LT(f_1),\dots,\LT(f_s)) \subseteq (\LT(f_1'),\dots,\LT(f_{s'}'))$.
            
                    \State $L_0 = \AscOrdLT(F), \pi_0 = \prod\limits_{\ell \in L_0} \LT(\ell), i=0$,
                    \While{True}
                    \State $L_{prec} = L_i, \pi_{prec} = \pi_i$
                    \State $L_{succ} = []$
                    \For{$j = 1,\dots,\#L_{prec}$}
                    \State $f_j' = \multiVarRed(L_{prec}(j),L_{succ}, <_{\tau})$ \textcolor{magenta}{// Because we have used $\AscOrdLT$ on $L_{prec}$, \\ \quad \quad \quad \quad \quad \quad  \quad \quad \quad \quad \quad \quad \quad \quad \quad \quad \quad \quad  \quad \quad \quad only the polynomials in the list with smaller \\ \quad \quad \quad \quad \quad \quad  \quad \quad \quad \quad \quad \quad \quad \quad \quad \quad \quad \quad  \quad \quad \quad  leading term can be used to divide}
                    \If{$f_j' \not= 0$}
                    \State $L_{succ} := L_{succ} \cup [f_j']$
                    \EndIf
                    \EndFor
                    \State $\pi_{succ} = \prod\limits_{\ell \in L_{succ}} \LT(\ell)$
                    \If{$\pi_{prec} = \pi_{succ}$}
                    \State $F' := \AscOrdLT(L_{succ})$
                    \State \textbf{break}
                    \EndIf
                    \State $i := i + 1$
                    \State $L_i = \AscOrdLT(L_{succ}), \pi_{i} = \pi_{succ}$
                    \EndWhile
                    \For{$j = 1,\cdots,\#F'$} \textcolor{magenta}{// We make the output monic}
                    \State $F'(j) := \frac{F'(j)}{\LC(F'(j))}$
                    \EndFor 
                    \State\Return $F'$
		\end{algorithmic}
\end{algorithm}

\begin{proposition}\label{prop:onestepinterred0}
Let $F = [f_1,\dots,f_s] \subseteq R$ be a list of non-zero pair-wise distinct polynomials and $\tau$ a term order. For any $i \in\{ 1,\dots,s\}$, let either $f_i' = \multiVarDiv(f_i, F \smallsetminus \{f_i\}, <_{\tau})$ or $f_i' = \multiVarRed(f_i, F \smallsetminus \{f_i\}, <_{\tau})$. Then, the following facts hold:
\begin{enumerate}[(1)]
\item $(F) = (F \smallsetminus \{f_i\}) + (f_i')$.
\item $(\LT(f_1),\dots,\LT(f_s)) \subseteq (\LT(f_1),\dots,\LT(f_{i-1}),\LT(f_i'),\LT(f_{i+1}),\dots,\LT(f_s))$, with the convention that $\LT(f_i') := 0$ if $f_i' = 0$.
\end{enumerate}
\end{proposition}
\begin{proof}
Let $i \in \{1,\dots,s\}$. If $f_i' = 0$, then the claim is trivially verified. So, we assume that $f_i' \not= 0$. Since $f_i'$ is the remainder of the division between $f_i$ and $F \smallsetminus \{f_i\}$,  we have that $f_i' - f_i \in (F \smallsetminus \{f_i\})$. Thus, we get that $(F) = (F \smallsetminus \{f_i\}) + (f_i')$, which proves \textit{(1)}. Moreover, since the operation in either $\multiVarDiv$ or $\multiVarRed$ is a polynomial division, we obtain that either $\LT(f_i) = \LT(f_i')$ or there exists an index $ j \in \{1,\dots,s\} \smallsetminus \{i\}$ such that $\LT(f_j) | \LT(f_i)$. In both cases, we have
$$\LT(f_i) \in (\LT(f_1),\dots,\LT(f_{i-1}),\LT(f_i'),\LT(f_{i+1}),\dots,\LT(f_s)),$$
and thus \textit{(2)} follows.
\end{proof}

\begin{theorem}
 Let $F = [f_1,\dots,f_s] \subseteq R$ be a list of non-zero pair-wise distinct polynomials, $\tau$ a term order, and let $F'=\interred(F,<_{\tau})$ the output of Algorithm~\ref{interredltfirst}. Then, $F'$ is a list of interreduced monic polynomials such that $(F')=(F)$ and $(\LT(F))\subseteq(\LT(F'))$.
\end{theorem}
\begin{proof}
The proof is divided into three parts: the first one tackles the termination of the algorithm, the second addresses its correctness and the third one concerns the last part of the statement.

We now address the termination. Suppose, by contradiction that the algorithm does not terminate. Then, the \textbf{while} loop goes on indefinitely. So, for all $i \in \mathbb{N}$ we have that at iteration $i$ of the \textbf{while} loop $\pi_{prec} \not= \pi_{succ}$. When we do $\multiVarDiv$ only three things can happen: $0 = f_j' \not = L_{prec}(j)$, $0 \not = f_j' = L_{prec}(j)$ or $0 \not= f_j' \not= L_{prec}(j)$. If $f_j' = 0$, the contribute of $\LT(L_{prec}(j))$ in $\pi_{succ}$ is gone; if $f_j' = L_{prec}(j)$, the contribute of $\LT(f_j')$ in $\pi_{succ}$ is the same one of the contribute of $\LT(L_{prec}(j))$ in $\pi_{prec}$; if $0 \not= f_j' \not= L_{prec}(j)$, since we are dealing with a polynomial division ($\multiVarDiv$), we have that $\LT(f_j') \leq_{\tau} \LT(L_{prec}(j))$; so $\pi_{prec} \geq_{\tau} \pi_{succ}$. But then, supposing that the $\textbf{while}$ loop does not terminate, we get that, at each iteration $i$ of the loop, $\pi_{prec} \not= \pi_{succ}$, so $\pi_{prec} >_{\tau} \pi_{succ}$. But, then, we have built an infinite descending sequence of terms, contradicting Lemma~\ref{descending chains}. $\lightning$ 

The correctness of the $\interred$ algorithm can be proved in the following way. Once we exit the \textbf{while} loop, we are sure that, for every $j \in \{1,\dots,\#L\}$, $\LT(L(j)) \not\in (\LT(L \smallsetminus [L(j)]))$, so there can not be other reductions between the leading terms: at the last iteration of the \textbf{while} loop, for $j = 1,\dots,\#L_{prec}$ we have that $\LT(f_j')$ is always equal to $\LT(L_{prec}(j))$, and this means that all the divisions $\multiVarDiv$ in this last iteration are done for the tails, therefore completing the interreduction.

The facts that $(F') = (F)$ and $(\LT(F)) \subseteq (\LT(F'))$ are immediate from Proposition~\ref{prop:onestepinterred0}, noting that all the operations done in the algorithm are reductions to $0$ and polynomial divisions.
\end{proof}

\begin{remark}
In the interreduction algorithm $\interred$, we could also have reduced all the leading terms in the \textbf{while} loop, applying $\multiVarRed$ instead of $\multiVarDiv$. Then after the end of the loop, we could have reduced the tails using $\multiVarDiv$. Termination and correctness of the algorithm can be proven also in this case, with analogous ideas from the ones we have given with our choice.
\end{remark}

\subsection{Buchberger's Algorithm}
From now on, we fix a term order $<$ on $R$.
How do we find a Gr\"obner basis of an ideal $I$? The first algorithm for doing that was introduced by Buchberger in his Ph.D.\ thesis (see \cite{Buch70, Buch06}).

\begin{proposition}\label{GB_and_division_with_remainder_prop}
    Let $I$ be an ideal of $R$, and let $g_1,\ldots,g_c \in I$. The following facts are equivalent:
    \begin{enumerate}
        \item The set $\left\{g_1,\ldots,g_c\right\}$ is a Gr\"{o}bner basis of $I$.
        \item For all $f\in I$ and for all divisions with remainder $f=q_1g_1+\dots+q_cg_c+r$, then $r=0$.
        \item For all $f\in I$ there exists a division with remainder $f=q_1g_1+\dots+q_cg_c+r$, with $r=0$.
    \end{enumerate}
\end{proposition}
\begin{proof}
\begin{itemize}
    \item[$1. \Rightarrow 2.$] Let $f \in I $ and consider a division with remainder
    $$f=q_1g_1+\dots+q_cg_c+r.$$
    Suppose by contradiction $r \not = 0$. Then, we still have that $\LT_<(r) \in \LT_<(I)$, so there exists $g \in \{g_1,\dots,g_c\}$ such that $\LT_<(g)\mid\LT_<(r)$. This constradicts Definition~\ref{division_with_remainder_def}. $\lightning$
    \item[$2. \Rightarrow 3.$] It is trivial.
    \item[$3. \Rightarrow 1.$] Let $f \in I\setminus\{0\}$. We have to prove that there exists $g \in \{g_1,\dots,g_c\}$ such that $\LT_<(g)\mid\LT_<(f)$. By Definition \ref{division_with_remainder_def}, we have that $\LT_<(f) \geq \LT_<(q_ig_i)$ for all $i = 1,\dots,c$. Thus, there exists $j \in \{1,\dots,c\}$ such that $\LT_<(f) = \LT_<(q_jg_j)$ (otherwise we cannot have the equality given in the division with remainder). So, we obtain
    $$\LT_<(f) = \LT_<(q_jg_j) = \LT_<(q_j)\LT_<(g_j),$$
    and then $\LT_<(g_j)\mid\LT_<(f)$.
    \end{itemize}
\end{proof}

Proposition~\ref{GB_and_division_with_remainder_prop} provides an equivalent characterization of a Gr\"obner basis. However, it is often impractical for direct application. In contrast, the criterion presented below serves as the foundation of Buchberger's Algorithm.
Before introducing this criterion, we must first provide the following definitions, which will also help us explain the F4 Algorithm in more detail.

\begin{definition}\label{Spoldef}\label{Shalfdef}
Let $f,g \in R$ be two non-zero polynomials. We define the \textbf{$S$-polynomial} of $f$ and $g$ as
$$S(f,g) = \sPol(f,g) = \frac{\LM(g)}{\gcd(\LT(f),\LT(g))}f-\frac{\LM(f)}{\gcd(\LT(f),\LT(g))}g.$$
We define the \textbf{$S$-half} of $f$ and $g$ as
$$\shalf(f,g) = \frac{\LM(g)}{\gcd(\LT(f),\LT(g))}f.$$
\end{definition}

The $S$-half of two non-zero polynomials $f,g \in R$ is just the minuend (the left part of the subtraction) in the $S$-polynomial $\sPol(f,g)$.
One can immediately verify that, given two non-zero polynomials $f,g \in R$ such that $f \not= g$ and (at least) one of them is not a monomial, we get that 
$$\shalf(f,g) \not= \shalf(g,f) \ \mbox{and} \ \sPol(f,g) = \shalf(f,g) - \shalf(g,f) \not= 0.$$

\begin{theorem}[\textbf{Buchberger's Criterion}]\label{BuchbergerCrit}
    Let $I=\left(g_1,\ldots,g_c\right) \subseteq R$ and assume that $g_1,\ldots,g_c$ are monic. Let $m_i := \LT_<(g_i)$. For $1 \le i < j \le c$ we denote the $S$-polynomial $\sPol(g_i, g_j)$ as $S_{i,j}$. The following facts are equivalent:
    \begin{enumerate}
        \item The set $\left\{g_1,\ldots,g_c\right\}$ is a Gr\"obner basis of $I$ w.r.t.\ $<$.
        \item For all $1 \le i < j \le c$ and for all division with remainder
        $S_{i,j} = q_1g_1+\dots+q_cg_c+r$, then $r=0$.        
        \item For all $1 \le i < j \le c$ there exists a division with remainder $S_{i,j} = q_1g_1+\dots+q_cg_c+r$ such that $r=0$. \end{enumerate}
\end{theorem}
\begin{proof}
 The implications \enquote{$1. \Rightarrow 2.$} and \enquote{$2. \Rightarrow 3.$} follow   from Proposition~\ref{GB_and_division_with_remainder_prop}. So, we are left to prove \enquote{$3. \Rightarrow 1.$}
   
 By assumption, for every $1 \le i < j \le c$ we have an expression
        \begin{equation}\label{BC1}
            S_{i,j} = q_{{i,j}_1}g_1 + \cdots + q_{{i,j}_c}g_c
        \end{equation}
        with $\LT_<(q_{{i,j}_k}g_k) \le \LT_<(S_{i,j})$ for all $k = 1,\dots,c$ such that $q_{{i,j}_k} \not= 0$. Recalling that
        \begin{equation}\label{BC2}
        S_{i,j} = \sPol(g_i, g_j) = \frac{m_j}{\gcd(m_i,m_j)}g_i-\frac{m_i}{\gcd(m_i,m_j)}g_j,
        \end{equation}
        we have that
        \begin{equation}\label{BC3}
             \LT_<(q_{{i,j}_k}g_k) \le \LT_<(S_{i,j}) \le \frac{m_i m_j}{\gcd(m_i,m_j)} = \lcm(m_i,m_j)
        \end{equation}
        for all $k = 1,\dots,c$ such that $q_{{i,j}_k} \not= 0$. 
    Now, substituting \eqref{BC2} into \eqref{BC1}, we get
    \begin{equation}\label{BC4}
        \frac{m_j}{\gcd(m_i,m_j)}g_i-\frac{m_i}{\gcd(m_i,m_j)}g_j - q_{{i,j}_1}g_1 - \cdots - q_{{i,j}_c}g_c = 0.
    \end{equation}
    Take $f \in I$. Thanks to Proposition~\ref{GB_and_division_with_remainder_prop}, to conclude we have to show that there exists a division of $f$ via $g_1,\dots,g_c$ with remainder $r = 0$. Since $f \in I$, we know that there exists an expression of the form
    \begin{equation}\label{BC5}
        f = h_1g_1 + \cdots + h_cg_c.        
    \end{equation}
    If $\LT_<(f) \ge \LT_<(h_ig_i)$ for all $i = 1,\ldots,c$ such that $h_c \not= 0$, this is a division with remainder $r = 0$ and we are done. If that is not the case, we have to \textit{twist} the equation and transform it into a division with remainder. The strategy is to do it step by step, showing that at each step the expression we get is better. We will associate to this construction a decreasing sequence of terms, and so the process ends with the desired result after a finite amount of steps. Let us take as guidance
    $h=(h_1,\ldots,h_c)$,
    $$T(h) := \max\{\LT_<(h_ig_i) \ | \ h_i \not= 0\}$$
    and
    $$A(h) := \{i \ | \ \LT_<(h_ig_i) = T\}.$$
    Since $\LT_<(f) < \LT_<(h_ig_i)$ for at least one $i$ (otherwise we are done), we have that $\LT_<(f) < T(h)$, and so $|A(h)| \geq 2$, otherwise $T(h)$ would survive in $h_1g_1 + \cdots + h_cg_c$.
    So, there exist $i,j \in A(h)$. From that we get
    $$\LT_<(h_ig_i) = \LT_<(h_jg_j),$$
    i.e.
    $$\LT_<(h_i) m_i = \LT_<(h_j) m_j.$$
    Dividing by $\gcd(m_i,m_j)$ we have that there exists a term $t \in \mathbb{T}$ such that
    $$\LT_<(h_i) = t \frac{m_j}{\gcd(m_i,m_j)} \ \mbox{and} \ \LT_<(h_j) = t \frac{m_i}{\gcd(m_i,m_j)}.$$
    Now, we add the left hand side of \eqref{BC4} (which is zero) multiplied by $\lambda t$ (for some $\lambda \in \kk \setminus \{0\}$) to the right hand side of \eqref{BC5} and we get
    $$f = h_1 g_1 + \cdots + h_c g_c + \lambda t \left(\frac{m_j}{\gcd(m_i,m_j)}g_i - \frac{m_i}{\gcd(m_i,m_j)}g_j - q_{{i,j}_1}g_1 - \cdots - q_{{i,j}_c}g_c\right),$$
    which can be rewritten as
    $$f = h^\prime_1 g_1 + \cdots + h^\prime_c g_c$$
    where
    \[h^\prime_k = h_k + \lambda t\cdot \begin{cases}
       \left( \frac{m_j}{\gcd(m_i,m_j)} - q_{{i,j}_i} \right) &\mbox{if $k=i$;}\\
        \left(-\frac{m_i}{\gcd(m_i,m_j)} - q_{{i,j}_j} \right) &\mbox{if $k = j$;}\\
        \left(- q_{{i,j}_k} \right) &\mbox{if $k \not= i,j$.}
    \end{cases}\]
    Here, we take $\lambda := - \LC_<(h_i)$; in this way we have that $\LT_<(h^\prime_i) < \LT_<(h_i)$ and, by denoting $h^\prime=(h^\prime_1,\ldots,h^\prime_c)$, one can verify that
    \[\mbox{$[T(h^\prime) < T(h)]$ or $[T(h^\prime) = T(h)$ and $A(h^\prime) \subsetneq A(h)]$.}\]
    
    The process stops when we get a division with remainder that we want. Suppose by contradiction that we do not get the desired output in a finite amount of steps. Then, since the cardinality of $A(h)$ is always smaller or equal than $c$, we can write an infinite descending sequence of terms of the following form
    $$T(h) > T(h^{(1)}) > \cdots > T(h^{(\ell)}) > \cdots,$$
    and this is a contradiction by Lemma~\ref{descending chains}. $\lightning$
\end{proof}

We are now ready to present a basic version of the Buchberger's Algorithm.
\begin{algorithm}[H]
		\caption{Buchberger's Algorithm}\label{Buchberger}
		\begin{algorithmic}
			\Require $G = \left\{g_1,\ldots,g_c\right\}$ a set of generators of $I$
			\Ensure A Gr\"obner basis of $I$
                \State $\textup{Pairs} := G \times G$
                \State $m := c$
                \While{$\textup{Pairs} \neq \emptyset$}
                \State Choose $(h,k) \in \textup{Pairs}$ and erase it from $\textup{Pairs}$
                \State $g_{m+1} := \sPol(h,k)$
                \State $g_{m+1} := \multiVarDiv(g_{m+1}, G, <)$
                \If{$g_{m+1}\neq 0$}
                    \State $m := m+1$
                    \State $\textup{Pairs} := \textup{Pairs}\cup \left\{\left(g_i, g_m\right) \mid i \in \left\{1, \dots, m-1\right\}\right\}$
                    \State $G := G \cup \left\{g_m\right\}$
                    \EndIf
                \EndWhile           
            \State\Return $G$
		\end{algorithmic}
\end{algorithm}

\textbf{Why does this procedure end?} \\
If $g_{c+1}\neq 0$ then $\LT_<(g_{c+1}) \notin J_0 := \left(\LT_<(g_1),\ldots,\LT_<(g_c)\right)$; so, if we set
$$J_1:= \left(\LT_<(g_1),\ldots,\LT_<(g_c), \LT_<(g_{c+1})\right),$$
we have the strict inclusion:
$$
J_0 \subsetneq J_1.
$$
Now, if $g_{c+2}\neq 0$ then $\LT_<(g_{c+2}) \notin J_1$, so if we define $J_2 := \left(\LT_<(g_1),\ldots, \LT_<(g_{c+1}), \LT_<(g_{c+2})\right)$, then again we have a strict inclusion:
$$
J_0 \subsetneq J_1 \subsetneq J_2
$$
Proceeding similarly we find out that the algorithm terminates, otherwise we obtain an increasing infinite chain of ideals in $R$, which is not possible since $R$ is Noetherian.

%We give a brief example of the Buchberger algorithm.

\begin{example}\label{guided example}
	Let $I=(f_1,f_2) \subseteq \mathbb{F}_{101}[x,y]$ with $f_1 = x^2+x+1$ and $f_2= xy-x$, and consider the $\DegRevLex$ term order. We compute a Gr\"{o}bner basis of $I$ using Buchberger's algorithm. Consider the $S$-polynomial
	
	$$S_{1,2} = yf_1 - xf_2 = x^2 + xy + y$$
	
	We perform the division of $S_{1,2}$ by $f_1$ and $f_2$:
	
	\begin{align*}
		\begin{array}{ccccc|cc}
			x^2 & + xy & & + y & & x^2+x+1, & xy-x\\
			\cline{6-7}
			\rule[0mm]{0cm}{5mm}
			x^2 & & +x & & +1 & 1 &  \\
			\cline{1-5}
			\rule[0mm]{0cm}{5mm}
			//  & xy & - x & +y & -1 & &1  \\ 
			& xy & -x & & & & \\
			\cline{2-5}
			\rule[0mm]{0cm}{5mm}
			& // & //  & y & -1 & & 
		\end{array}	
	\end{align*}
	
	Thus, we obtain $S_{1,2} = f_1 + f_2 + r$ with $r=y-1$.
	Let $f_3:= r$ and add it to the list of generators. At this point, observe that $f_3$ is a divisor of $f_2$, so we can eliminate $f_2$ from our set of generators. We apply Buchberger's criterion to $\{f_1, f_3\}$; we need to test the $S$-polynomial $S_{1,3}$. However, it is easy to check that $S_{1,3}$ has remainder $0$ when divided by $f_1$ and $f_3$ since $  \LT_{\DRL}(f_1)$ and $\LT_{\DRL}(f_3)$ are coprime (in fact, $S_{1,3} = f_1 + (x+1)f_3$). Therefore, $\{f_1, f_3\}$ is the desired Gr\"{o}bner basis. Observe that it is already in reduced form.
\end{example}

\subsection{Polynomial System Solving}\label{sec system resolution}
We now go back to the PoSSo Problem mentioned in the introduction, that is to solve a multivariate polynomial system $f_1=\cdots=f_c=0$ as~\eqref{sistema}. We can consider the ideal $I = \left(f_1, \dots, f_c\right) \subseteq R$ associated to the system $F=\{f_1,\dots, f_c\}$.
The {\bf zero locus} of $I$ is the set

$$
Z(I) = \left\{P\in \overline{\kk}^n \,\mid \, \forall f \in I \ f(P) = 0 \right\} 
= \left\{P \in \overline{\kk}^n \, \mid \, \forall i  \in \{1,\ldots,r \} \ f_i(P)=0 \right\} .
$$

\noindent Solving the polynomial system~\eqref{sistema} is equivalent to finding the zero locus $Z(I)$. Notice that the zero locus of $I$ consists of the solutions of the polynomial system over the algebraic closure $\overline{\kk}$ of $\kk$.
If we are interested in the solutions over the field $\kk$, then we may want to restrict our attention to $Z(I)\cap\kk^n$.

The main connection between polynomial system solving and Gr\"obner bases is established through the $\Lex$ term order. In essence, once a lexicographic Gr\"obner basis is available, computing the zero locus of the corresponding ideal becomes computationally straightforward. This is typically best explained using the Shape Lemma (Theorem~\ref{thm:shapelemma}). Before stating the result, we recall three notions that serve as assumptions. We will later discuss whether they are truly necessary.

\begin{remark}
Note that a polynomial system has no solution (over $\overline{\kk}$) if and only if $I = R$. Testing if a polynomial system has solution, is equivalent to test whether $1\in I$. This can be done by looking at any Gr\"obner basis $G$ of $I$: the element $1$ is in $I$ if and only if $G$ contains a constant element.
\end{remark}

\begin{theorem}\label{0-dim tfe}
Let $I \subseteq R$ be an ideal, then the following facts are equivalent:
\begin{enumerate}
\item $|Z(I)|< +\infty$.
\item $\dim_\kk(R/I) < +\infty$.
\item For all $i \in \{1, \dots, n\}$, $I \cap \kk[x_i] = (h_i(x_i))$ with $h_i \neq 0$.
\end{enumerate}
\end{theorem}
\begin{definition}
If the equivalent conditions of Theorem~\ref{0-dim tfe} are satisfied, we say that $I$ is {\bf zero-dimensional} and we call the \textbf{degree} of the ideal $I$ the quantity $\deg(I):= \dim_\kk(R/I)$.    
\end{definition}

\begin{definition}
    The ideal $I$ is said to be in {\bf normal form} with respect to the variable $x_n$ if for all $P=(p_1,\ldots,p_n),\ Q = (q_1,\ldots,q_n) \in Z(I)$ with $P \neq Q$, we have $p_n \ne q_n$.
\end{definition}

\begin{definition}
    The {\bf radical ideal} of $I$ is the ideal 
    $$
        \sqrt{I} = \left\{ f \in R \, \mid \, \exists n \in \NN^*  \ \mbox{s.t.} \ f^n \in I \right\}
    $$
    The ideal $I$ is called {\bf radical} if $I = \sqrt{I}$ .
\end{definition}

\begin{comment}
\begin{example}
    $R = \kk[x]$ \\
    $I = ((x-1)^2)$ \text{is not radical} because $(x-1) \in \sqrt{I} \setminus I$ \\
    \text{In this case} $\sqrt{I} = (x-1)$   
\end{example}
\end{comment}
\begin{theorem}[\textbf{Shape Lemma}]\label{thm:shapelemma}
    Let $\kk$ be a perfect field, let $I \subseteq R$ be an ideal which is radical, zero--dimensional and in normal form with respect to $x_n$. Then, the reduced $\Lex$ Gr\"{o}bner basis of $I$ has the form:
    $$
        \left\{x_1-g_1(x_n),x_2-g_2(x_n),\ldots,x_{n-1}-g_{n-1}(x_n),g_n(x_n)\right\}, 
    $$
    where $\deg(g_n)=|Z(I)|.$
\end{theorem}

The Shape Lemma gives us a way to easily recover the elements of $Z(I)$ from the $\Lex$ Gr\"obner basis: we factorize $g_n$ to find its roots, then to each root of $g_n$ corresponds a point in $Z(I)$.
Consequently, if univariate polynomial factorization is efficient over $\kk$ (for instance, if $\kk$ is a finite field), then having a lexicographic Gr\"obner basis allows us to efficiently solve a polynomial system.
However, the Shape Lemma relies on certain assumptions. While we focus on zero-dimensional ideals—since characterizing an infinite zero locus can be challenging—we argue that the other two hypotheses, normal form with respect to $x_n$ and radicality, are not strictly necessary. We discuss this in the next remark. For a more thorough exposition, see, for example, \cite{CG21}.

\begin{remark}\label{rem:assumptionshape} Consider the set up of Theorem~\ref{thm:shapelemma}.
\begin{itemize}
\item \enquote{$I$ in normal form with respect to $x_n$}. Every zero--dimensional ideal can be brought in normal form with respect to $x_n$ through a linear coordinate change as explained e.g. in \cite[Proposition~3.7.22]{KR1}. Notice however, that if the field is finite, a field extension might also be needed. 
\item \enquote{$I$ radical}. If the field is perfect, there are efficient algorithms for computing the radical of $I$ such as \cite[Corollary~3.7.16]{KR1}. Since $Z(I) = Z\left(\sqrt{I}\right)$, it is sufficient then to apply the Shape Lemma to to the radical of $I$. 
\end{itemize}
\end{remark}
The strategy outlined in Remark~\ref{rem:assumptionshape} for handling ideals that do not satisfy the assumptions of the Shape Lemma may not be optimal. In fact, the lexicographic Gr\"obner basis of a zero--dimensional, non radical ideal often retains a useful \textit{shape} that enables straightforward computation of its zero locus. This property is, in essence, a consequence of the Elimination Theorem (see \cite[Chapter~3]{CLS97}). We formalize this observation in the following theorem.

\begin{theorem}\label{thm:lexbasis}
    Let $I$ be a proper and zero--dimensional ideal of $R$. The reduced lexicographic Gr\"{o}bner basis of $I$ has the form:
    $$
    \begin{aligned}
        & p_{n, 1}\left(x_n\right), \\
        & p_{n-1,1}\left(x_{n-1}, x_n\right), \ldots, p_{n-1, t_{n-1}}\left(x_{n-1}, x_n\right), \\
        & p_{n-2,1}\left(x_{n-2}, x_{n-1}, x_n\right), \ldots, p_{n-2, t_{n-2}}\left(x_{n-2}, x_{n-1}, x_n\right), \\
        & \vdots \\
        & p_{1,1}\left(x_1, \ldots, x_n\right), \ldots, p_{1, t_1}\left(x_1, \ldots, x_n\right),
    \end{aligned}
    $$
    where $p_{i, t_j} \in R$ for every $i \in\{1, \ldots, n\}, j \in\left\{1, \ldots, t_i\right\}$ and $t_1, \ldots, t_{n-1} \geq 1$. Moreover, for any $1 \leq \ell \leq n$, let $a=\left(a_{\ell+1}, \ldots, a_n\right) \in \kk^{n-\ell}$ be a solution of the equations:
    $$
    \begin{aligned}
        & p_{n, 1}\left(x_n\right), \\
        & p_{n-1,1}\left(x_{n-1}, x_n\right), \ldots, p_{n-1, t_{n-1}}\left(x_{n-1}, x_n\right), \\
        & \vdots \\
        & p_{\ell+1,1}\left(x_{\ell+1}, \ldots, x_n\right), \ldots, p_{\ell+1, t_{\ell+1}}\left(x_{\ell+1}, \ldots, x_n\right),
    \end{aligned}
    $$
    and let
    $$	p_{\ell}\left(x_{\ell}\right)=\gcd\left(p_{\ell, 1}\left(x_{\ell}, a_{\ell+1}, \ldots, a_n\right), \ldots, p_{\ell, t_{\ell}}\left(x_{\ell}, a_{\ell+1}, \ldots, a_n\right)\right) .
    $$
    Then $p_{\ell}\left(x_{\ell}\right) \notin \kk$.
\end{theorem}

Theorem~\ref{thm:lexbasis} provides a strategy for efficiently computing the zero locus of $I$. The process begins by factoring the univariate polynomial  $p_{n,1}(x_n)$ and substituting its roots into the remaining polynomials, thereby obtaining new univariate polynomials in $x_{n-1}$ and so on. For further details, we refer the reader to \cite{CG21}.

\subsection{Macaulay Matrices}\label{sec Macaulay matrices}
Other algorithms for computing Gr\"obner bases have been developed after Buchberger's (see e.g. \cite{MutantXL, XLalgorithm, F4paper, Fau02}). In many of these, the concept of the Macaulay matrix plays a crucial role. This is the case for the F4 Algorithm which we will explain in Section~\ref{sec F4}.
In fact, some of these algorithms build on Lazard's idea \cite{Laz83} of converting the computation of a Gr\"obner basis into multiple instances of Gaussian elimination. As before, we fix a polynomial ring $R = \kk[x_1,\dots,x_n]$ over a field $\kk$.

\begin{definition}
	Let $F = \left\{f_1, \dots, f_c\right\} \subseteq R$, let $\tau$ be a term order on $R$, and let $d \in\NN^*$. Let $\mathbb{T}_{\leq d}$ be the set of terms in $R$ of degree less than or equal to $d$. The \textbf{Macaulay matrix} $\mathcal{M}_{\leq d}$ of $F$ is a matrix whose columns are indexed by the terms in $\mathbb{T}_{\leq d}$ ordered with respect to $\tau$, and whose rows are indexed by the polynomials $t_{h,k}f_k$, where $k = 1, \dots, c$, $h = 1, \dots, n_k$ for some integer $n_k$, and with $t_{h,k}\in \mathbb{T}_{\leq d}$ such that $\deg(t_{h,k}f_k) \leq d$. The $(i,j)$ entry of $\mathcal{M}_{\leq d}$ is the coefficient of the $j$-th term in the polynomial of the $i$-th row.

	If $F = \left(f_1, \dots, f_c\right)$ is a set of homogeneous generators, we consider instead the \textbf{homogeneous Macaulay matrix} $\mathcal{M}_d$, which is constructed similarly to the matrix $\mathcal{M}_{\leq d}$, but whose columns are indexed by the terms of degree exactly $d$, and whose rows are indexed by the polynomials $t_{h,k}f_k$, where $k = 1, \dots, c$, $h = 1, \dots, n_k$ for some integer $n_k$, and  with $t_{h,k}\in \mathbb{T}_{\leq d}$ such that $\deg(t_{h,k}f_k) = d$.

\end{definition}

We also introduce the following notation, which will be useful to deal with submatrices of the Macaulay matrix.

\begin{definition}
Let $A = (a_{i,j}) \in \mbox{M}_{\ell,s}(\kk)$ be a $\ell \times s$ matrix with coefficients in $\kk$. Suppose that the $s$ columns of $A$ are indexed from left to right by a descending sequence of terms (of $R$) in $\TT$ of the form $t = [t_1 > t_2 > \cdots > t_s]$. Then, with $\rowsp_{t}(A)$, or, if there is no ambiguity $\rowsp(A)$, we denote the $\kk$-vector subspace of $R$ generated by the polynomials arising from the rows of $A$, i.e
$$\rowsp_{t}(A) = \langle a_{i,1}t_1 + a_{i,2}t_2 + \cdots + a_{i,s}t_s \mid i = 1,\dots,\ell\rangle_{\kk}.$$
Moreover, with $\mbox{rows}_{t}(A)$, or, if there is no ambiguity, with $\rows(A)$, we denote the set of the polynomials given by the rows of $A$ with respect to $t$, i.e.
$$\mbox{rows}_{t}(A) = \{a_{i,1}t_1 + a_{i,2}t_2 + \cdots + a_{i,s}t_s \ | \ i = 1,\dots,\ell\} \subseteq R.$$
Viceversa, if we have a list of polynomials $F = [f_1,\dots,f_{\ell}] \subseteq R$, and we consider the ordered list of terms $t = \Supp(F) = [t_1 > \cdots > t_s]$, we denote with $\mbox{matrix}(F)$ the $\ell \times s$ matrix $B = (b_{i,j})$ such that $b_{i,j}$ is $\mbox{coefficient of} \ t_j \ \mbox{in} \ f_i$:
$$\mbox{matrix}(F)_{i,j} = \mbox{coefficient of }\ t_j \ \mbox{in} \ f_i.$$
\end{definition}

It is easy to see that, if we have a list of polynomials $F = [f_1,\dots,f_{\ell}] \subseteq R$, if we take $t = \Supp(F)$, we get that 
$$\mbox{rows}_t(\mbox{matrix}(F)) = F.$$

Recall that in the Macaulay matrices $\mathcal{M}_{\leq d}$ of $F = [f_1,\dots,f_c]$ the columns are indexed by the list $t$ of the terms in $\mathbb{T}_{\leq d}$ ordered from the biggest to the smallest with respect to $\tau$. So, if we consider 
$$MF = [t_{h,k}f_k \ | \ k = 1,\dots,c, \ h = 1,\dots,n_k, \ \deg(t_{h,k}f_k) \leq d],$$
we have that $\asmatrix(MF)$ is a sub-matrix of $\mathcal{M}_{\leq d}.$ 

Let us see some examples of the previous definitions. 

\begin{example}
We fix $\tau$ as the term order DegLex on $R=\mathbb{F}_{101}[x,y]$ with $x {>_{\tau}} y$.
We have
$$ A:= \begin{pmatrix}
    1 & 21 & 63 \\
    100 &3 & 0 
\end{pmatrix} \in M_{2,3}(\mathbb{F}_{101}), \ m := [x^3 {>_{\tau}} x^2 y {>_{\tau}} y],$$
then
$$\mbox{rows}_m(A) = \{x^3 + 21 x^2 y + 63 y, 100x^3 + 3 x^2 y\}.$$
Vice versa, if $F = [f_1 = x^5 y + 5 x^2 y + 94 x, f_2 = x^5 y + 100y] \subseteq \mathbb{F}_{101}[x,y]$ and $m = \Supp(F) = [x^5 y {>_{\tau}} x^2 y {>_{\tau}} x {>_{\tau}} y]$ then
$$\mbox{matrix}(F) = \bordermatrix{
      & x^5y & x^2y & x & y \cr
    f_1 & 1 & 5 & 94 & 0 \cr
    f_2 & 1 & 0 & 0 & 100
}.$$
\end{example}

\begin{example}
   We fix $\tau$ as the term order DegLex on $R=\mathbb{F}_{101}[x,y]$ with $x {>_{\tau}} y$. Let $F= \left\{f_1, f_2 \right\}$ with $$f_1 = 3x+1 \text{ and } f_2 =xy+4y$$ and $d=3$.
    The matrix $\mathcal{M}_{\leq3}$ is given by:
	
	$$\bordermatrix{ & x^3 & x^2y & xy^2 & y^3 & x^2 & xy & y^2 & x & y & 1 \cr
		x^2f_1 & 3 & 0 & 0 & 0 & 1 & 0 & 0 & 0 & 0 & 0\cr
		xyf_1 & 0 & 3 & 0 & 0 & 0 & 1 & 0 & 0 & 0 & 0\cr
		y^2f_1 & 0 & 0 & 3 & 0 & 0 & 0 & 1 & 0 & 0 & 0\cr
		xf_1 & 0 & 0 & 0 & 0 & 3 & 0 & 0 & 1 & 0 & 0\cr
		yf_1 & 0 & 0 & 0 & 0 & 0 & 3 & 0 & 0 & 1 & 0\cr
		f_1 & 0 & 0 & 0 & 0 & 0 & 0 & 0 & 3 & 0 & 1\cr
		xf_2 & 0 & 1 & 0 & 0 & 0 & 4 & 0 & 0 & 0 & 0\cr
		yf_2 & 0 & 0 & 1 & 0 & 0 & 0 & 4 & 0 & 0 & 0\cr
		f_2 & 0 & 0 & 0 & 0 & 0 & 1 & 0 & 0 & 4 & 0 }$$	
\end{example}
 In the next example we compute the degree reverse lexicographic Gr\"obner basis of an ideal $I$ using Macaulay matrices.
\begin{comment}
 \begin{example}
Let's consider the homogeneous system $\mcF = \left\{f_1,f_2\right\} \subseteq \QQ[x,y,z]$ where 
$$
    f_1 = x^2+2xy+xz \,\,\text{ and }\,\, f_2 = xy+yz+z^2 .
$$
The Macaulay matrix $\mathcal{M}_{3}$ of $\mcF$ in degree $3$ is given by:	
	$$\bordermatrix{ & x^3 & x^2y & xy^2 & y^3 & x^2z & xyz & y^2z & xz^2 & yz & z^3 \cr
		xf_1 & 1 & 2 & 0 & 0 & 1 & 0 & 0 & 0 & 0 & 0\cr
		yf_1 & 0 & 1 & 2 & 0 & 0 & 1 & 0 & 0 & 0 & 0\cr
		xf_2 & 0 & 1 & 0 & 0 & 0 & 1 & 0 & 1 & 0 & 0\cr
		yf_2 & 0 & 0 & 1 & 0 & 0 & 0 & 1 & 0 & 1 & 0}$$
Reducing this matrix we obtain $\widetilde{\mathcal{M}}_{3}$:
$$\bordermatrix{ &x^3 & x^2y & xy^2 & y^3 & x^2z & xyz & y^2z & xz^2 & yz & z^3 \cr
		 &1 & 0 & 0 & 0 & 1 & -2 & 0 & -2 & 0 & 0\cr
		 &0 & 1 & 0 & 0 & 0 & 1 & 0 & 1 & 0 & 0\cr
	& 0 & 0 & 1 & 0 & 0 & 0 & 0 & -1/2 & 1 & 0\cr
	  & 0 & 0 & 0 & 0 & 0 & 0 & 1 & 1/2 & 0 & 0}.$$
Notice that the first row comes from $xf_1-2xf_2$. This reduction could have been done already in degree $2$!
If we consider $\mathcal{M}_{2}$ and we reduce it we obtain $\widetilde{\mathcal{M}}_{2}$:
$$\bordermatrix{ & x^2 & xy & y^2 & xz & yz & z^2 \cr
		 f_1-2f_2 & 1 & 0 & 0 & 1 & -2 & -2\cr
		 f_2 & 0 & 1 & 0 & 0 & 1 & 1 }.$$
The first row of $\widetilde{\mathcal{M}}_{3}$ is $x$ times the first row of $\widetilde{\mathcal{M}}_{2}$. This fact suggests us to consider the Macaulay matrices degree by degree and 
\end{example}
\end{comment} 

\begin{example}\label{esempio guida}
We fix $\tau$ as the term order DegLex on $R=\mathbb{F}_{101}[x,y]$ with $x {>_{\tau}} y$.
    Let $I=(f_1,f_2) \subseteq R$ where $f_1 = x^2+xy+x+1$ and $f_2= xy-x$.
  
   We consider first $\mathcal{M}_{\leq 2}$:
    $$\bordermatrix{ & x^2 & xy & y^2 & x & y & 1 \cr
        f_1 & 1 & 1 & 0 & 1 & 0 & 1\cr
        f_2 & 0 & 1 & 0 & -1 & 0 & 0 }.$$
    Reducing the matrix we obtain:
    $$\bordermatrix{ & x^2 & xy & y^2 & x & y & 1 \cr
        f_1-f_2 & 1 & 0 & 0 & 2 & 0 & 1\cr
        f_2     & 0 & 1 & 0 & -1 & 0 & 0 }.$$
    This matrix is reduced but $\{f_1-f_2, f_2\}$ is not a Gr\"{o}bner basis of $I$. Now, we consider $\mathcal{M}_{\leq 3}$:	
	$$\bordermatrix{ & x^3 & x^2y & xy^2 & y^3 & x^2 & xy & y^2 & x & y & 1 \cr
		xf_1 & 1 & 1 & 0 & 0 & 1 & 0 & 0 & 1 & 0 & 0\cr
		xf_2 & 0 & 1 & 0 & 0 & -1 & 0 & 0 & 0 & 0 & 0\cr
		yf_1 & 0 & 1 & 1 & 0 & 0 & 1 & 0 & 0 & 1 & 0\cr
		yf_2 & 0 & 0 & 1 & 0 & 0 & -1 & 0 & 0 & 0 & 0\cr
		f_1 & 0 & 0 & 0 & 0 & 1 & 1 & 0 & 1 & 0 & 1\cr
		f_2 & 0 & 0 & 0 & 0 & 0 & 1 & 0 & -1 & 0 & 0 }$$	
	We reduce the matrix and obtain:
	$$\bordermatrix{ & x^3 & x^2y & xy^2 & y^3 & x^2 & xy & y^2 & x & y & 1 \cr
		(x-2)f_1-(x-2)f_2 & 1 & 0 & 0 & 0 & 0 & 0 & 0 & -3 & 0 & -2\cr
		xf_2 & 0 & 1 & 0 & 0 & 0 & 0 & 0 & 2 & 0 & 1\cr
		yf_2+f_2 & 0 & 0 & 1 & 0 & 0 & 0 & 0 & -1 & 0 & 0\cr
		f_1-f_2 & 0 & 0 & 0 & 0 & 1 & 0 & 0 & 2 & 0 & 1\cr
		f_2 & 0 & 0 & 0 & 0 & 0 & 1 & 0 & -1 & 0 & 0\cr
		(y-1)f_1-(x+y+1)f_2& 0 & 0 & 0 & 0 & 0 & 0 & 0 & 0 & 1 & -1
		 }$$	
		 From there we obtain the Gr\"{o}bner basis $\{x^2+2x+1, y-1\}$.
\end{example}

\begin{comment}
questo è più vero per matrix F5!!!
From the example we can observe that in the second step, that is when considering the Macaulay matrix $\mathcal{M}_{\leq 3}$, we are doing again the reduction $f_1 - f_2$, already done in the first step. To overcome that, we can add the new polynomial $f_1 - f_2$ in the set of generators $\mcF$ on which we're building the Macaulay matrix. 
\end{comment}

The strategy used in Example~\ref{esempio guida} already embodies the core idea behind one of the simplest linear algebra-based algorithms for computing Gr\"obner bases---namely, the algorithm proposed by Lazard in \cite{Laz83}. More precisely, we outline the algorithm in the following remark.

\begin{remark}[Linear algebra based algorithms]\label{lazard_algorithm}
Let  $F\subseteq R$ be a finite set of polynomials. 
\begin{enumerate}
    \item Fix an integer $d >0$, build the Macaulay matrix $\mathcal{M}_{\leq d}$ of $F$ and perform Gaussian elimination on it to obtain a matrix in reduced row echelon form (RREF).
    \item Any row $\ell$ in the RREF of $\mathcal{M}_{\leq d}$  corresponds to a polynomial
$f_{\ell}$ of degree $\leq d$. If $\deg(f_{\ell}) < d$, we add new rows to $\mathcal{M}_{\leq d}$ corresponding to the polynomials $u f_{\ell}$ where $u$ is a term, $\deg(u f_{\ell}) \leq d$ and $u f_{\ell} \notin \text{rowsp}(\mathcal{M}_{\leq d})$.
\item Repeat the computation of the RREF and the operation of adding new rows, until there are no new rows to add. Denote by $\mathcal{M}^{MUT}_{\leq d}$ the matrix in RREF computed via this algorithm.
\end{enumerate}
It is clear that $\text{rowsp}(\mathcal{M}^{MUT}_{\leq d}) \subseteq (F)_{\leq d}$, where $(F)_{\leq d} = (F) \cap \{g \in R \ | \ \deg(g) \leq d\}$. For a given $d$, one may have $\text{rowsp}(\mathcal{M}^{MUT}_{\leq d}) \neq (F)_{\leq d}$. However, for $d \gg 0$, we have $\text{rowsp}(\mathcal{M}^{MUT}_{\leq d}) = (F)_{\leq d}$ and thus $\text{rowsp}(\mathcal{M}^{MUT}_{\leq d})$ contains a Gr\"{o}bner basis of $F$. On the other hand, if no termination criterion is provided (such as the one in Theorem~\ref{BuchbergerCrit}), it may be unclear when to stop the previous procedure. In Section~\ref{sec F4}, we will see how the F4 Algorithm overcomes this issue.
\end{remark}

%%%%%%%%%%%%%%%%%%%%%%%%%%%%%%%%%%%%%%%%%%%%%%%%%%%%%%%%%%%%%%%%%%%
%%%%%%%%%%%%%%%%%%%%%%%%%  Sezione %%%%%%%%%%%%%%%%%%%%%%	

%%%%%%%%%%%%%%%%%%%%%%%%%%%%%%%%%%%%%%%%%%%%%%%%%%%%%%%%%%%%%%%%%%%
%%%%%%%%%%%%%%%%%%%%%%%%%  Sezione %%%%%%%%%%%%%%%%%%%%%%	
\newpage

\section{FGLM Algorithm}\label{sec FGLM}
The FGLM algorithm was introduced in 1993 by Faugére, Gianni, Lazard, and Mora in \cite{FGLM}. 
It is a linear-algebra-based algorithm that allows to change the ordering of a given Gr\"{o}bner basis of a zero-dimensional ideal. More precisely, given $I \subseteq R = \kk[x_1, \dots, x_n]$ a zero-dimensional ideal and  two distinct term orders on $R$, the
algorithm takes in input a reduced Gr\"{o}bner basis of $I$ w.r.t.\ one term order and gives in output a reduced Gr\"{o}bner basis of $I$ w.r.t.\ the other term order.
Let $D := \deg(I)$, we will see that the complexity of the algorithm is bounded from above by $O(n\cdot D^3)$ field operations.
Note that the zero-dimensional assumption is crucial for the algorithm to work. \\

\textbf{Nota Bene.}
{In this chapter $I$ will always be a zero-dimensional ideal of degree $D$.} We will also assume that $Z(I) \neq \emptyset$ to avoid the trivial case where $I = R$ and simplify the writing of the algorithms.\\

Suppose  that we know a Gr\"obner basis $G$ of the ideal $I$ with respect to a fixed term order $<$, then, thanks to a theorem of Macaulay (see \cite[Theorem 1.5.7]{KR1}), a standard basis for the $\kk$-vector space $R/I$  is given by:
\begin{align*}
     \left\{t \in \TT \, | \, \forall {g \in G}\ \LT_<(g) \nmid t\right\},
\end{align*}
where we say that a term $t$ is not (top) reducible by $G$ if $\LT_<(g) \nmid t$ for every $g\in G$.

We order the basis elements (ascending from left to right) with respect to $<$. In the following we will call it the \textbf{staircase} of $I$ and denote it by $\mathcal{W}_<(G)$, having
$$\mathcal{W}_<(G) = \left\{1 = w_1 < w_2 < \cdots < w_D\right\},$$
where $D$ is the dimension. Given $f\in R$,
we recall that the normal form $\NF_<(f,I)$ is the expression of $\overline{f} \in R/I$ with respect to the the basis $\mathcal{W}_<(G)$.

\subsection{The Idea of the Algorithm}
During the execution of the algorithm, we will compute several normal forms of some terms in a specific order. The polynomials of the new Gr\"{o}bner basis will arise as linear combinations of those. The same technique can be applied to compute, for $i=1,\ldots,n$, the univariate polynomials $p_i(x_i)$ such that $(p_i) = I \cap \kk[x_i]$ in $\kk[x_i]$ (the existence of $p_i\neq0$ is ensured by Theorem~\ref{0-dim tfe}). 
To give an idea of how the FGLM algorithm works, we now explain how to find $p_i$.
First, notice that, since $R/I$ has dimension $D$, the elements 
$$\overline{1}, \, \overline{x_i}, \, \overline{x_i^2}, \dots, \overline{x_i^D}$$ are surely linearly dependent.
So we can proceed as follows. We start from $\NF_<(1,I) = 1$ and we compute recursively $$\NF_<(x_i^j, I) = \NF_<\left(x_i\cdot\NF_<\left(x_i^{j-1}, I\right), I\right) \,\, \text{ for } j = 1, \dots, D .$$
At each step we write those normal forms as vectors $v_0,v_1,\ldots$ in the basis $\mathcal{W}_<(G)$, where 
$v_0 = [1,0,\dots,0]$,
and proceed until we can find a non-trivial linear dependency among the vectors.
When we find a linear combination $$0 = \lambda_0v_0 + \dots + \lambda_dv_d \text{ for } d\leq D,$$
we obtain a polynomial $$\lambda_0 + \lambda_1x_i + \dots + \lambda_dx_i^d \in I$$
that has minimal degree and hence coincides (up to a constant factor) with $p_i$.

\subsection{The Property of the Staircase and the Border}
Let us start with an example.
\begin{example}\label{1st ex staircase}
Let $\kk = \FF_{101}$, $R = \kk[x,y]$ with the $\DegRevLex$ order and $I = \left( xy^5-x^2, xy^2-y\right) \subseteq R$.
The reduced Gr\"obner basis of $I$ is
$$G = \left\{xy^2-y, y^4-x^2, x^3-y^3\right\} .$$
It follows that the $\kk$-vector space $R/I$ has dimension $8$ and the staircase of $G$ is given by 
$$\mathcal{W}_<(G) = \left\{1 < y < x < y^2 < xy < x^2 < y^3 < x^2y\right\}.$$
We can represent the staircase in a $xy$-graph where a point $(i,j)$ corresponds to the monomial $x^iy^j$:
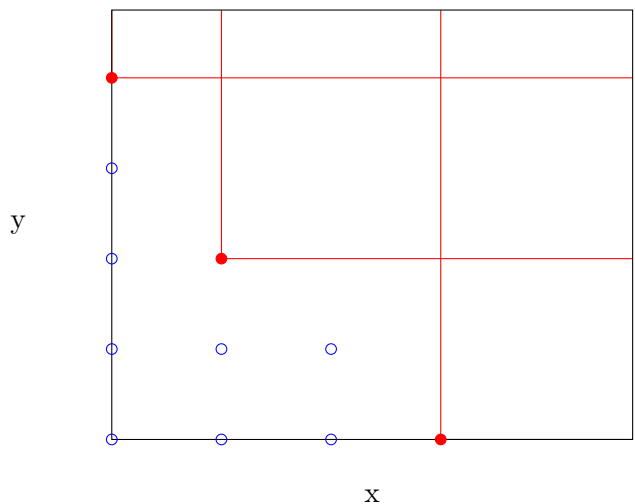
\begin{center}
    % \vspace*{0.6cm}
%\resizebox{0.5\linewidth}{!}{
\begin{tikzpicture}
[scale=1, line cap=round ,
% Styles
%axes/.style=,
 axes/.style={
%     axis background/.style={fill=white},  % white or transparent background?
     xlabel=$\mbox{x}$, ylabel=$\mbox{y}$,
%     axis x line=middle, axis y line=center,
%     axis equal=true,
%     grid=both, %minor, % major, none
   },
important line/.style={very thick},
information text/.style={rounded corners,fill=red!10,inner sep=1ex}]
\newcommand{\PiX}{3}
\newcommand{\PiY}{1}
\newcommand{\PiiX}{2}
\newcommand{\PiiY}{4}
\newcommand{\XMax}{4}
\newcommand{\XMaxx}{4.75}
\newcommand{\YMax}{4}
\newcommand{\YMaxx}{4.75}
\begin{axis}[ ymin=0, ymax=\YMaxx, xmin=0, xmax=\XMaxx,
  xlabel={x}, ylabel={y}, ylabel style={rotate=-90},
  axis on top,
  xtick=\empty, ytick=\empty]
  % empty dots
%%  \foreach \n in {0,1,...,\XMax}
%%  \foreach \m in {0,1,...,\YMax}
%%  \addplot[mark=o]  coordinates {(\n,\m)};
  \addplot[blue, mark=o]  coordinates {(0,0)};
  \addplot[blue, mark=o]  coordinates {(0,1)};
  \addplot[blue, mark=o]  coordinates {(1,0)};
  \addplot[blue, mark=o]  coordinates {(0,2)};
  \addplot[blue, mark=o]  coordinates {(1,1)};
  \addplot[blue, mark=o]  coordinates {(2,0)};
  \addplot[blue, mark=o]  coordinates {(2,1)};
  \addplot[blue, mark=o]  coordinates {(0,3)};
  % black dots
  \addplot[red, mark=*]  coordinates {(1,2)};
  \addplot[red, mark=*]  coordinates {(0,4)};
  \addplot[red, mark=*]  coordinates {(3,0)};
  \addplot[red, domain=1:\XMaxx, samples=2]  (x,2);
  \addplot[red, domain=2:\YMaxx, samples=2]  (1,x);
  \addplot[red, domain=0:\XMaxx, samples=2]  (x,4);
  \addplot[red, domain=4:\YMaxx, samples=2]  (0,x);
  \addplot[red, domain=3:\XMaxx, samples=2]  (x,0);
  \addplot[red, domain=0:\YMaxx, samples=2]  (3,x);
  %\addplot[red, mark=o]  coordinates {(1,4)};
  %\addplot[red, mark=o]  coordinates {(2,2)};
  %\addplot[red, mark=o]  coordinates {(2,3)};
\end{axis}
\end{tikzpicture}
%}%end resizebox
\end{center}

Here the blue empty balls are the elements of the staircase and the red balls are the leading terms of the elements of the Gr\"obner basis. All the other terms that are not shown are elements of $\LT_<(I)$. We can see from this graph that  the elements of $\mathcal{W}_<(G)$ indeed form a staircase!
\end{example}

We see as a straightforward consequence of the definition of staircase, that $\mathcal{W}_<(G)$ is closed under division. More precisely, we have the following.
\begin{proposition}
Let $\omega \in \mathcal{W}_<(G) \setminus \left\{1\right\}$ and let $t \in \TT$ such that $t \mid \omega$, then $t \in \mathcal{W}_<(G)$.
\end{proposition}
\begin{proof}
    By contradiction, let $\omega \in \mathcal{W}_<(G) \setminus \left\{1\right\}$ and let $t \in \TT$ such that $t \mid \omega$ and $t \not\in \mathcal{W}_<(G)$. Then, by definition of staircase, there exists $g \in G$ such that $\LT(g) \mid t$. But, then, $\LT(g) \mid w$, and, again using the definition of staircase, this is a contradiction. $\lightning$
\end{proof}

\begin{definition}
We define the \textbf{border of the staircase} as the set:
$$\mcF_<(G) := \left\{x_i \omega \, \middle| \, \omega \in \mathcal{W}_<(G), 1 \leq i \leq n \text{ and } x_i\omega \notin \mathcal{W}_<(G)\right\} .$$
\end{definition}

\begin{example}
We continue Example \ref{1st ex staircase}.
The border of the staircase is given by:
$$\mcF_<(G) := \left\{xy^2,  x^3, y^4, xy^3, x^2y^2, x^3y\right\} .$$
In the following graph, the red balls (empty or full) represent the elements of the border.
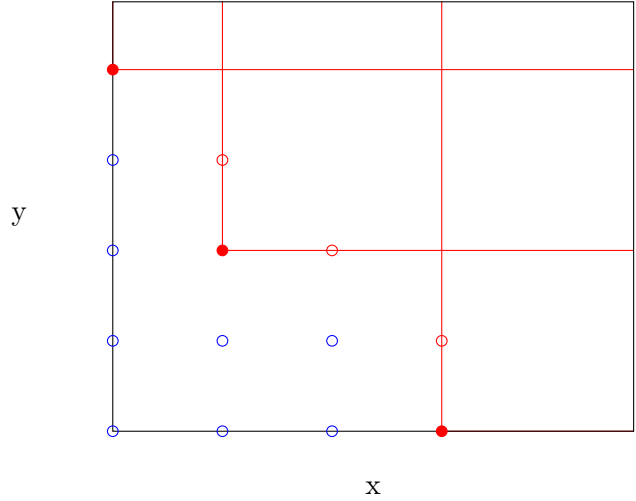
\begin{center}
    % \vspace*{0.6cm}
% \centering\resizebox{0.75\linewidth}{!}{
\begin{tikzpicture}
[scale=1, line cap=round ,
% Styles
%axes/.style=,
 axes/.style={
%     axis background/.style={fill=white},  % white or transparent background?
     xlabel=$x$, ylabel=$y$,
%     axis x line=middle, axis y line=center,
%     axis equal=true,
%     grid=both, %minor, % major, none
   },
important line/.style={very thick},
information text/.style={rounded corners,fill=red!10,inner sep=1ex}]
\newcommand{\PiX}{3}
\newcommand{\PiY}{1}
\newcommand{\PiiX}{2}
\newcommand{\PiiY}{4}
\newcommand{\XMax}{4}
\newcommand{\XMaxx}{4.75}
\newcommand{\YMax}{4}
\newcommand{\YMaxx}{4.75}
\begin{axis}[ ymin=0, ymax=\YMaxx, xmin=0, xmax=\XMaxx,
  xlabel={x}, ylabel={y}, ylabel style={rotate=-90},
  axis on top,
  xtick=\empty, ytick=\empty]
  % empty dots
%%  \foreach \n in {0,1,...,\XMax}
%%  \foreach \m in {0,1,...,\YMax}
%%  \addplot[mark=o]  coordinates {(\n,\m)};
  \addplot[blue, mark=o]  coordinates {(0,0)};
  \addplot[blue, mark=o]  coordinates {(0,1)};
  \addplot[blue, mark=o]  coordinates {(1,0)};
  \addplot[blue, mark=o]  coordinates {(0,2)};
  \addplot[blue, mark=o]  coordinates {(1,1)};
  \addplot[blue, mark=o]  coordinates {(2,0)};
  \addplot[blue, mark=o]  coordinates {(2,1)};
  \addplot[blue, mark=o]  coordinates {(0,3)};
  % black dots
  \addplot[red, mark=*]  coordinates {(1,2)};
  \addplot[red, mark=*]  coordinates {(0,4)};
  \addplot[red, mark=*]  coordinates {(3,0)};
  \addplot[red, domain=1:\XMaxx, samples=2]  (x,2);
  \addplot[red, domain=2:\YMaxx, samples=2]  (1,x);
  \addplot[red, domain=0:\XMaxx, samples=2]  (x,4);
  \addplot[red, domain=4:\YMaxx, samples=2]  (0,x);
  \addplot[red, domain=3:\XMaxx, samples=2]  (x,0);
  \addplot[red, domain=0:\YMaxx, samples=2]  (3,x);
  \addplot[red, mark=o]  coordinates {(1,3)};
  \addplot[red, mark=o]  coordinates {(2,2)};
  \addplot[red, mark=o]  coordinates {(3,1)};
\end{axis}
\end{tikzpicture}
% }%end resizebox
\end{center}

\end{example}
From the previous example we can note that the elements of $\LT_<\left(G\right)$ are the minimal elements of $\mcF_<(G)$ with respect to the partial order of division. This is not a coincidence, but a general fact.
\begin{proposition}
    $\LT_<\left(G\right) \subseteq \mcF_<(G)$.
\end{proposition}
\begin{proof}
Let $g \in G$; then for every $i \in \left\{1, \dots, n\right\}$ such that $x_i \mid \LT_<(g)$ we have that $\dfrac{\LT_<(g)}{x_i} \notin \LT_<(I)$ and so $\dfrac{\LT_<(g)}{x_i} \in  \mathcal{W}_<(G)$. By definition we have $\LT_<(g) = x_i\dfrac{\LT_<(g)}{x_i} \in \mcF_<(G)$.
\end{proof}

\begin{proposition}\label{minimality of LTG}
Let $G$ be a reduced Gr\"obner basis of $I$, then for each $t \in \mcF_<(G)$ we have two possible cases:
\begin{enumerate}
\item There exists $g \in G$ such that $t = \LT_<(g)$;
\item There exists $j \in \left\{1, \dots, n\right\}$ and $t^\prime \in \mcF_<(G)$ such that $t = x_jt^\prime$.
\end{enumerate}
\end{proposition}
\begin{proof}
Let $t \in \mcF_<(G)$ and set $A_t := \left\{j \in \left\{1, \dots,n\right\} \, \middle| \, x_j \mid t \text{ and } \dfrac{t}{x_j} \notin \mathcal{W}_<(G)\right\}$.

First, assume that $A_t$ is empty. Since $t \notin \mathcal{W}_<(G)$, then $t \in \LT_<(G)$ and so $t = \LT_<(g)u$ for some $g \in G$ and $u \in \TT$. We show that $u = 1$. If $x_j \mid u$ for some $j \in \left\{1, \dots, n\right\}$ then $\dfrac{t}{x_j} = \LT_<(g)\dfrac{u}{x_j} \notin \mathcal{W}_<(G)$ and hence $j \in A_t$ that is a contradiction. So $u = 1$ and $t = \LT_<(g)$.

Now, assume that $A_t$ is not empty. Then, there exists $j \in \left\{1, \dots, n\right\}$ such that $x_j \mid t$ and $t^\prime = \dfrac{t}{x_j} \notin \mathcal{W}_<(G)$. Since $t \in \mcF_<(G)$, there exists $\omega \in \mathcal{W}_<(G)$ and $i \in \left\{1, \dots,n\right\}$ such that $t = x_i\omega$. Note that $i \neq j$ since $t^\prime \notin \mathcal{W}_<(G)$, so $x_j \mid \omega$ and $\omega^\prime = \dfrac{\omega}{x_j} \in \mathcal{W}_<(G)$. This means that $t^\prime = x_i \omega^\prime \in \mcF_<(G)$. Since $t = t^\prime x_j$, the proof is completed.
\end{proof}

\begin{remark}\label{Gbound}
It follows from the definition of $\mcF_<(G)$, recalling that $\left|\mathcal{W}_<(G)\right|=D$, that $\left| \mcF_<(G)\right| \leq nD$. Since $\LT_<\left(G\right) \subseteq \mcF_<(G)$ we have a bound on the numbers of elements of a reduced Gr\"obner basis of $I$. More precisely, if $G$ is a reduced Gr\"obner basis of $I$ we have
$$\left| G\right| \leq nD .$$
We will use this bound for the complexity analysis of the FGLM algorithm in Theorem~\ref{thm:complexityFGLM}.
\end{remark}

\subsection{A First Description of the FGLM Algorithm}
We can now give a first non-detailed description of the algorithm.
Recall that, given $I\subseteq R$ a zero-dimensional ideal and  two distinct term orders on $R$, namely $<_1$ and $<_2$,  the algorithm takes in input a reduced Gr\"{o}bner basis of $I$ w.r.t.\  $<_1$ and gives in output a reduced Gr\"{o}bner basis of $I$ w.r.t.\  $<_2$.
The algorithm will test the terms of $\TT$ in increasing order, building the new staircase $S$ and the new Gr\"obner basis $G$. For each tested term we want to understand whether it is in the staircase $S$ or in $\LT_{<_2}(G)$; all the terms that are multiples of elements of $\LT_{<_2}(G)$ will not be tested.

In the following pseudo-code, to simplify the notation, we will write $\NF_1(f)$ instead of $\NF_{<_1}\left(f,I\right)$ for $f \in R$. Moreover, given an object $l$ of length $d$ (for example, a vector or a list), we indicates its entries with $l_1, \dots, l_d$.
\begin{algorithm}[H]\label{first FGLM}
		\caption{FGLM algorithm}
		\begin{algorithmic}
			\Require $I \neq R$ an ideal, $\mathcal{G}_1$ a reduced Gr\"{o}bner basis of $I$ w.r.t.\ $<_1$ and $<_2$ a new term order
			\Ensure $\mathcal{G}_2$ a reduced Gr\"{o}bner basis of $I$ w.r.t.\ $<_2$			
			\State $L := \text{Sort}([X_1, \dots, X_n], <_2)$ \quad \textcolor{magenta}{//ordered by $<_2$} 		
			\State $S:= [1]$ \quad \textcolor{magenta}{//the new staircase under construction for $<_2$}
                \State $V := [1]$ \quad \textcolor{magenta}{//the $\NF_1$ of $S$}			
			\State $G:= []$ \quad \textcolor{magenta}{//the new G.b.} 
			\While{$L \neq \emptyset$}
				\State $t:= \text{First}(L)$ \quad \textcolor{magenta}{//$t = \min_{<_2}(L)$}	
                \State $L := L\setminus \left[t\right]$
				\State $v := \NF_1(t)$  \quad \textcolor{magenta}{// $\mbox{NF}_1(t) = \multiVarDiv(t, \mathcal{G}_1, <_1)$}
                \State $r := \#S$
				\If{$v \in \left\langle V\right\rangle_{\kk}$} 
					\State find $(\lambda_i)_{i=1}^r$ such that $v = \sum_{i=1}^{r}\lambda_iV_i$
					\State $G := G \cup  \left[t-\displaystyle\sum_{i=1}^{r}\lambda_iS_i\right]$
				\Else
					\State $S := S \cup \left[t\right]$       \State $V := V \cup \left[v\right]$
					\State $L := \text{Sort}(L \cup \left[X_it \mid i \in \left\{1,\dots,n\right\}\right], <_2)$
					\State Remove doubles and multiples of $\LT_{<_2}\left(G\right)$ from $L$
				\EndIf
			\EndWhile						
			\State\Return $G$
		\end{algorithmic}
\end{algorithm}
 
We now give a brief description of the algorithm. It may be not very clear why this algorithm works and why the \enquote{while} loop ends, but we will give a proof of the correctness of the algorithm later in Theorem \ref{FGLM correctness}. Since we are assuming $I \neq R$, we know that $1 \in S$, so we initialize $S$ as the list of  length one containing the term $1$. The list $V$ will contain the normal forms with respect to the first term order of the elements of $S$, so we initialize it as well as the  list containing the term $1$.
The list $L$ will contain at each step the terms that need to be tested. Its elements will always be in increasing order w.r.t.\ $<_2$. At the beginning of the algorithm, $L$ is the list containing all the variables. At each step of the algorithm we test the first element $t$ of $L$, erasing it from the list and we compute its normal form $v$. If $v$ is already in the vector space generated by the elements of $V$, this means that $t$ is an element of $\LT_{<_2}(I)$ and more precisely, it is an element of $\LT_{<_2}\left(G\right)$. From the linear combination that we found, we can find the correspondent polynomial in $G$. Otherwise, $t$ is an element of the staircase, so we add it to $S$. We also add to $L$ all the terms obtained by multiplying $t$ with a variable. At the end of each step we erase doubles from $L$ and multiples of elements already in $\LT_{<_2}\left(G\right)$. 

The fact that the algorithm works is due to the good properties of the staircase and the border that we saw in the previous subsection. Since the staircase is closed under division, we can proceed by multiples until we test an element of the border: thanks to Proposition \ref{minimality of LTG}, among the latter, we will first test the elements of $\LT_{<_2}\left(G\right)$.

Now, we want to give a matricial and effective version of the previous algorithm: how do we compute the normal forms in an efficient way? How do we test linear dependency?
We will answer these two questions in the next subsections.

\subsection{Multiplication Matrices}
During the algorithm, we will compute several times the normal form $\NF_1(x_k p)$ where $p$ is a polynomial already in its normal form and $x_k$ is a variable. We see in the following that we can easily do this computation as a matrix and vector multiplication. 
Suppose now that we fixed a term order $<$ and the staircase $\mathcal{W}_<(G)$.
\begin{definition}
Let $$\Phi_k : \, R/I \longrightarrow R/I, \quad \overline{f} \mapsto \overline{x_k f}.$$
The map $\Phi_k$ is a linear map between $\kk$-vector spaces, so we can consider its representation matrix.
We define the $\bf{k}$\textbf{-th multiplication matrix} as the representation matrix $\bf{M^{(k)}}$ of the map $\Phi_k$ in the standard basis $\mathcal{W}_<(G)$.
\end{definition}
If $\mathcal{W}_<(G) = \left\{1 = w_1 < w_2 < \cdots < w_D\right\}$, then the matrix $\bf{M^{(k)}}$ is a $D \times D$ matrix such that
$$M_{i,j}^{(k)} = \text{the coefficient of } w_i \text{ in } x_kw_j, \quad \text{for } i,j \in \left\{1, \dots, D\right\}$$
\begin{example}\label{esempio guida multmat}
    Consider $\mathcal{G}_{DRL} := \left\{y^2 + 34 x + y + 2, x^2 + xy+2y \right\} \subseteq \mathbb{F}_{101}\left[x,y\right]$, one can check that this is a Gr\"obner basis for a zero-dimensional ideal $I$ w.r.t.\ $\DegRevLex$. We have that $\deg(I) = 4$ and:
    
    $$\mathcal{W}(\mathcal{G}_{DRL}) = \left\{1 <_{DRL} y <_{DRL} x <_{DRL} xy\right\} = \left\{w_1, w_2, w_3, w_4\right\}.$$ 
    We now want to compute the multiplication matrices $M^{(1)}$ and $M^{(2)}$.
    We have: 
\begin{align*}
xw_1 &= x = w_3\\
xw_2 &= xy = w_4\\
\NF(xw_3) &= -2w_2-w_4\\
\NF(xw_4) &= 4w_1 + 35w_2-31w_3-33w_4\\
\end{align*}
and so:
$$M^{(1)} = \bordermatrix{ & xw_1 & xw_2 & xw_3 & xw_4 \cr
    w_1 & 0 & 0 & 0 & 4\cr
    w_2 & 0 & 0 & -2 & 35 \cr
    w_3 & 1 & 0 & 0 & -31 \cr
    w_4 & 0 & 1 & -1 & -33 } .$$    
Similarly we can compute:
\begin{align*}
yw_1 &= y = w_2\\
\NF(yw_2) & = -2w_1-w_2-34w_3\\
yw_3 &= xy = w_4\\
\NF(yw_4) &= -33w_2-2w_3+33w_4\\
\end{align*}
and obtain:
    $$M^{(2)} = \bordermatrix{ & yw_1 & yw_2 & yw_3 & yw_4 \cr
    w_1 & 0 & -2 & 0 & 0\cr
    w_2 & 1 & -1 & 0 & -33 \cr
    w_3 & 0 & -34 & 0 & -2 \cr
    w_4 & 0 & 0 & 1 & 33 }.$$
\end{example}

We note from the last example that a term can appear as a column of two different multiplication matrices (in the example $xw_2 = yw_3$). We also have various divisibility relations among the terms. These remarks suggests that there is a more efficient way to compute the multiplication matrices all together.

\subsection{The Transition Matrix}
To give an effective version of the algorithm, we want to detect linear dependency using a $D \times D$ matrix. 
For this purpose, we want to build during the algorithm the change of basis matrix $P$ of $R/I$ from the old basis $\mathcal{W}_{<_1}(\mathcal{G}_1)$ to the new basis $\mathcal{W}_{<_2}(\mathcal{G}_2)$ under construction.
To do that, we initialize $P$ as the identity matrix $I_D$. Suppose now that during the algorithm we have: $$S = \left\{s_1, \dots, s_r\right\} \text{ and } V = \left\{v_1, \dots, v_r\right\},$$
where now the $v_i$'s are written as vectors in the basis $\mathcal{W}_{<_1}(\mathcal{G}_1)$.
At this step we want $P$ to be such that for all $i \in \left\{1, \dots, r\right\}$:
$$Pv_i = e_i,$$
where \[
\begin{matrix}
e_{i} = &
        &(0,\ldots,0,&1,&0,\ldots,0)^T .\\
        &&&i \\
\end{matrix}
\]
Suppose that during the algorithm we find $v = \NF_{1}(t)$ and we want to test linear dependency between $v$ and $v_1, \dots, v_r$. To do that we compute
$$\lambda = \left(\lambda_1, \dots, \lambda_D \right) = Pv .$$
If for all $i \in \left\{r+1, \dots, D\right\}$ we have $\lambda_i = 0$, then clearly $v - \displaystyle\sum_{i=1}^r \lambda_iv_i = 0$ and so during the algorithm we find the new polynomial of the Gr\"obner basis $$t - \displaystyle\sum_{i=1}^r \lambda_is_i .$$
Otherwise, if there exists $i \in \left\{r+1, \dots, D\right\}$ such that $\lambda_i \neq 0$, then $v, v_1, \dots, v_r$ are linearly independent: in the latter case we add $t$ to $S$, $v$ to $V$ and we update $P$ to a new $D \times D$ invertible matrix $P^\prime$ such that
$$P^\prime v_i = e_i \text{ for } i \in \left\{1, \dots, r\right\} \text{ and } P^\prime v = e_{r+1} .$$
How do we find $P^\prime$? We now give a procedure to update the matrix $P$ using the vector $\lambda$.
\begin{comment}
$r+1$ linear indipendent vectors $v_1, \dots, v_r, v_{r+1}$ such that $P \cdot v_i = e_i$ for $i \in \left\{1, \dots, r\right\}$ 
a new $D\times D$ invertible matrix $P'$ such that $P'v_i = e_i$ for $i \in \left\{1, \dots, r+1\right\}$ 
\end{comment}

\begin{algorithm}[H]
    \caption{The Update procedure}\label{Update}
    \begin{algorithmic}
        \Require a $D\times D$ invertible matrix $P$, $r \in \left\{1, \dots D\right\}$ and a vector $\lambda = Pv$
        \Ensure a new $D\times D$ invertible matrix $P'$ such that $P'v = e_{r+1}$		
        \State $\lambda := P  v$ 
        \State $k := \min\left\{i \in \left\{r+1, \dots, D\right\}\mid \lambda_i \neq 0\right\}$
        \State $P[r+1, \ast] \longleftrightarrow P[k, \ast]$
        \State $\lambda[r+1] \longleftrightarrow \lambda[k]$
        \State $P[r+1, \ast] := P[r+1, \ast]/\lambda[r+1]$
        \For{$i =1, \dots, D, i \neq r+1$}
         \State $P[i,\ast] := P[i, \ast] - P[r+1,\ast]\cdot\lambda[i]$
        \EndFor
        \State\Return $P$
    \end{algorithmic}
\end{algorithm}
We now give a proof that this algorithm actually does what we need.
\begin{lemma}\label{correctness of Update}
Let $P \in \GL_D(\kk)$ and let $v_1, \ldots, v_r, v$ be $r+1$ linear independent vectors such that for all $i \in \left\{1, \dots, r\right\}$, $P v_i = e_i$, where $e_i$ is the $i$-th basis vector. Let $\lambda=P v$, then after the procedure of Algorithm~\ref{Update} , the matrix $P$ remains invertible and satisfies $P v_i=e_i$ for $i \in\{1, \ldots, r\}$ and $P v=e_{r+1}$.
\end{lemma}
\begin{proof}
Assume first that $\lambda_{r+1} \neq 0$. Then the output of Algorithm~\ref{Update} corresponds to the left multiplication of $P$ by the matrix $T=\left(t_{i, j}\right)_{1 \leq i, j \leq D} \in \operatorname{M}_D(\kk)$ with
$$
t_{r+1, r+1}=1 / \lambda_{r+1}, \quad t_{i, r+1}=-\lambda_i / \lambda_{r+1} \text { and } t_{i, i}=1 \text { for } i \neq r+1, \quad t_{i, j}=0 \text { otherwhise. }
$$
Since $T$ is invertible, $P$ remains invertible after the procedure. Moreover, $T e_i=e_i$ for $1 \leq i \leq r$, hence the property $T P v_i=e_i$ for $i \leq r$ remains unchanged. Lastly, we can note that $TPv=T \lambda=e_{r+1}$.\\
Now, if $\lambda_{r+1}=0$, the procedure looks for the smallest $k>r+1$ such that $\lambda_k \neq 0$ (which exists since $v \notin \left<v_1, \ldots, v_r\right>_\kk$ means $\lambda =P v \notin \left<e_1, \ldots, e_r\right>_\kk$ ) and swaps the $k$-th and the $(r+1)$-th rows of $P$ and $\lambda$. All assumptions on $P$ and $\lambda$ are kept, but now $\lambda_{r+1} \neq 0$ and we can conclude thanks to the first part of the proof.
\end{proof}

\subsection{The Matricial Version of FGLM}
We are now ready to give the effective version of  the FGLM Algorithm.
\begin{algorithm}[H]
    \caption{FGLM algorithm}\label{matricial FGLM}
    \begin{algorithmic}
        \Require $\mathcal{G}_1$ a reduced Gr\"{o}bner basis of $I\neq R$ w.r.t.\ $<_1$, $<_2$ a new term order and $M^{(1)}, \dots, M^{(n)}$
        \Ensure $\mathcal{G}_2$ a reduced Gr\"{o}bner basis of $I$ w.r.t.\ $<_2$			
        \State $L := \text{Sort}([(i,1) \mid i \in \left\{1, \dots, n\right\}], <_2)$ \quad\textcolor{magenta}{//list of pairs $(k,l)$ which correspond to $X_k \cdot S_l$ ordered by $<_2$} 		
        \State $S:= [1]$ \quad\textcolor{magenta}{//the staircase for $<_2$} 
        \State $V := [(1,0,\dots,0)^T]$ \quad\textcolor{magenta}{//the $NF_1$ of $S$ written in the basis $\mathcal{W}(\mathcal{G}_1)$}
        \State $G:= []$ \quad\textcolor{magenta}{//the new G.b.} 
        \State $P := I_D$ \quad\textcolor{magenta}{//the transition matrix}
        \While{$L \neq \emptyset$}
        \State $t:= \text{First}(L) = (k,l)$ 
        \State $L := L\setminus \left[t\right]$
        \State $r := \#S$
        \State $v := M^{(k)}V_l$ \quad\textcolor{magenta}{//$v$ is the $NF_1$ of $X_k\cdot S_l$ written in the basis $\mathcal{W}(\mathcal{G}_1)$}
        \State $\lambda := P  v$ \quad\textcolor{magenta}{//$\lambda$ detects linear dependencies}
        \If{$\lambda_{r+1} = \dots = \lambda_D = 0$}
        \State $G := G \cup  \left[X_k \cdot S_l-\displaystyle\sum_{i=1}^{r}\lambda_iS_i\right]$
        \Else
        \State $P:=$Update$(P,r,\lambda)$
        \State $S := S \cup \left[X_k \cdot S_l\right]$, $V := V \cup \left[v\right]$
        \State $L := \text{Sort}(L \cup \left[(i,r) \mid i \in \left\{1,\dots,n\right\}\right], <_2)$
        \State Remove doubles and multiples of $LT_{<_2}(G)$ from $L$
        \EndIf
        \EndWhile        
        \State\Return $G$
    \end{algorithmic}
\end{algorithm}

\subsection{A Toy Example}
To show how to apply the FGLM algorithm, we continue from Example~\ref{esempio guida multmat}, where we have
$\mathcal{G}_{DRL} := \left\{y^2 + 34 x + y + 2, x^2 + xy+2y \right\} \subseteq \mathbb{F}_{101}\left[x,y\right]$, $D = \deg(I) = 4$ and
	$$\mathcal{W}(\mathcal{G}_{\DRL}) = \left\{1 <_{\DRL} y <_{\DRL} x <_{\DRL} xy\right\} = \left\{w_1, w_2, w_3, w_4\right\}.$$ 
We want to find the $\Lex$ reduced Gr\"{o}bner basis, so we will apply the FGLM algorithm with $<_1 = \DRL$ and $<_2 = \Lex$. We already computed the multiplication matrices:

\begin{minipage}{.5\linewidth}%
    $$M^{(1)} = \bordermatrix{ & xw_1 & xw_2 & xw_3 & xw_4 \cr
        w_1 & 0 & 0 & 0 & 4\cr
        w_2 & 0 & 0 & -2 & 35 \cr
        w_3 & 1 & 0 & 0 & -31 \cr
        w_4 & 0 & 1 & -1 & -33 }$$
\end{minipage}%
\begin{minipage}{.5\linewidth}%
    $$M^{(2)} = \bordermatrix{ & yw_1 & yw_2 & yw_3 & yw_4 \cr
        w_1 & 0 & -2 & 0 & 0\cr
        w_2 & 1 & -1 & 0 & -33 \cr
        w_3 & 0 & -34 & 0 & -2 \cr
        w_4 & 0 & 0 & 1 & 33 }$$
\end{minipage}%
\vspace*{0.4cm}

\textbf{Setup}: $L = [(2,1)\textcolor{magenta}{= y},(1,1)\textcolor{magenta}{= x}]$, $S = [1]$, $V = [(1,0,0,0)^T\textcolor{magenta}{= \omega_1}]$, $G = []$, $P = I_4$.
\begin{enumerate}[Step 1:]
\item We start with $t = (2,1) \textcolor{magenta}{= y}$. We compute
    $v = M^{(2)} \cdot V_1 = (0,1,0,0)^T \textcolor{magenta}{= \omega_2}$ and
    $\lambda = P \cdot v = v$. Since $\lambda_2 \neq 0$, then:
    $$S := [1, y] \text{ and } V := [\omega_1, \omega_2] .$$
    The procedure Update on the matrix $P$ does nothing in this case.
    We remove $t$ from $L$, add its multiples and sort w.r.t.\ $\Lex$:
    $$L := \left[(2,2)\textcolor{magenta}{= y^2}, (1,1)\textcolor{magenta}{= x}, (1,2)\textcolor{magenta}{= xy}\right] .$$
\item We have $t = (2,2) \textcolor{magenta}{= y^2}$. We compute
    $v = M^{(2)} \cdot V_2 = (-2,-1,-34,0)^T \textcolor{magenta}{= \omega_3}$ and
    $\lambda = P \cdot v = v$. Since $\lambda_3 \neq 0$, then:
    $$S := [1, y, y^2] \text{ and } V := [\omega_1, \omega_2, \omega_3] .$$
    The procedure Update on the matrix $P$ gives us:
    $$
    P := \begin{pmatrix}
			1 & 0 & -6 & 0\\
			0 & 1 & -3 & 0 \\
			0 & 0 & -3 & 0 \\
			0 & 0 & 0 & 1 
		\end{pmatrix}
    $$
    We remove $t$ from $L$, add its multiples and sort w.r.t.\ $\Lex$:
    $$L := \left[(2,3)\textcolor{magenta}{= y^3}, (1,1)\textcolor{magenta}{= x}, (1,2)\textcolor{magenta}{= xy}, (1,3)\textcolor{magenta}{= xy^2}\right] .$$
\item We have $t = (2,3) \textcolor{magenta}{= y^3}$. We compute
$v = M^{(2)} \cdot V_3 = (2,-1,34,-34)^T \textcolor{magenta}{= \omega_4}$ and
$\lambda = P \cdot v = (0,-2,-1,-34)^T$. Since $\lambda_4 \neq 0$, then:
$$S := [1, y, y^2, y^3] \text{ and } V := [\omega_1, \omega_2, \omega_3, \omega_4] .$$
The procedure Update on the matrix $P$ gives us:
$$
P := \begin{pmatrix}
        1 & 0 & -6 & 0\\
        0 & 1 & -3 & -6 \\
        0 & 0 & -3 & -3 \\
        0 & 0 & 0 & -3
    \end{pmatrix}
$$
We remove $t$ from $L$, add its multiples and sort w.r.t.\ $\Lex$:
$$L := \left[(2,4)\textcolor{magenta}{= y^4}, (1,1)\textcolor{magenta}{= x}, (1,2)\textcolor{magenta}{= xy}, (1,3)\textcolor{magenta}{= xy^2}, (1,4)\textcolor{magenta}{= xy^3}\right] .$$
\item We have $t = (2,4) \textcolor{magenta}{= y^4}$. We compute
$v = M^{(2)} \cdot V_4 = (2,14,1,23)^T $ and
$\lambda = P \cdot v = (-4,-26,29,32)^T$. We have just found a polynomial of the new Gr\"obner basis:
$$
    G = [y^4-32y^3-29y^2+26y+4] .
$$
$$L := \left[(1,1)\textcolor{magenta}{= x}, (1,2)\textcolor{magenta}{= xy}, (1,3)\textcolor{magenta}{= xy^2}, (1,4)\textcolor{magenta}{= xy^3}\right] .$$
\item We have $t = (1,1) \textcolor{magenta}{= x}$. We compute
$v = M^{(1)} \cdot V_1 = (0,0,1,0)^T $ and
$\lambda = P \cdot v = (-6,-3,-3,0)^T$. We've just found a polynomial of the new Gr\"obner basis:
$$
    G = [y^4-32y^3-29y^2+26y+4, x+3y^2+3y+6] .
$$
All the elements of $L$ are multiples of $x$ so we empty $L$ and the algorithm terminates.
\end{enumerate}

\subsection{Proof of Correctness and  Complexity Analysis}
We now prove the correctness of Algorithm~\ref{matricial FGLM}.
\begin{theorem}\label{FGLM correctness}
Algorithm~\ref{matricial FGLM} finishes and computes a Gr\"obner basis of $I$ w.r.t.\ $<_2$.
\end{theorem}
\begin{proof}
First, we prove that, at the end of the algorithm, we have $S = \mathcal{W}_{<_2}(\mathcal{G}_2)$.
	
\noindent Let $\mathcal{W}_{<_2}(\mathcal{G}_2) = \left\{\omega_1, \dots, \omega_D\right\}$ and for $i = 1, \dots, D$ let $S^i = \left\{\sigma_1, \dots, \sigma_i\right\}$ be the state of the list $S$ when we added its $i$-th element during the algorithm. Suppose $S \neq \mathcal{W}_{<_2}(\mathcal{G}_2)$ and let %\textcolor{green}
{$k = \min\left\{i \mid S^i \neq \left\{\omega_1, \dots, \omega_i \right\} \right\}$}. Note that $k >1$ since $S^1 = \left\{1\right\} = \left\{\omega_1\right\}$. There are three cases:
\begin{itemize}
    \item $S^k = \left\{\sigma_1, \dots, \sigma_{k-1}\right\}$ so the algorithm stops with $\#S = k-1. $
    	
    Let $x_j \mid \omega_k$ then $\dfrac{\omega_k}{x_j} \in \left\{\omega_1, \dots, \omega_{k-1}\right\} = \left\{\sigma_1, \dots, \sigma_{k-1}\right\}$, so we tested $\omega_k$ and did not add it to $S$. This means that  $\left\{\sigma_1, \dots, \sigma_{k-1}, \omega_k\right\}$ is linear dependent, but $\left\{\sigma_1, \dots, \sigma_{k-1}, \omega_k\right\} = \left\{\omega_1, \dots, \omega_k\right\}$ and we find a contradiction.
    \item $\omega_k <_2 \sigma_k$:
    
    We can proceed exactly as in the previous case to find a contradiction.
    \item $\sigma_k <_2 \omega_k$:
    
    $\sigma_k \notin \mathcal{W}_{<_2}(\mathcal{G}_2)$ so $\sigma_k \in LT_{<_2}(I)$ and $0 = \sigma_k - \displaystyle\sum_{i=1}^{k-1}a_i\omega_i = \sigma_k - \displaystyle\sum_{i=1}^{k-1}a_i\sigma_i$ with $a_i \in \kk$, that is a contradiction since $\left\{\sigma_1, \dots, \sigma_k\right\}$ is a linear independent set.
\end{itemize}	
We now prove that at the end of the algorithm we have %\textcolor{red}
{$G = \mathcal{G}_2$.}
The fact that $G \subseteq \mathcal{G}_2$ is clear.
Let now $g \in \mathcal{G}_2$, then for all $x_j$ such that $x_j \mid \LT_2(g)$, we have $\dfrac{\LT_2(g)}{x_j} \in \mathcal{W}_{<_2}(\mathcal{G}_2)$; this means that we added $\LT_2(g)$ to $L$ at some point of the algorithm. Moreover for all $t \in \Supp(g) \setminus\left\{\LT_2(g)\right\}$, $t = \omega_{j_t} \in \mathcal{W}_{<_2}(\mathcal{G}_2)$. Let $l := \max \left\{j_t \mid t \in \Supp(g) \setminus\left\{LT(g)\right\} \right\}$, then when during the algorithm we test $\LT_2(g)$ we find that $\omega_1, \dots, \omega_l, \LT_2(g)$ are linear dependent, and $\NF_1(\LT(g)) = \displaystyle\sum_{i = 1}^l \lambda_i\omega_i$, so we add the polynomial $\LT(g) - \displaystyle\sum_{i = 1}^l \lambda_i\omega_i = g$ in $G$.
\end{proof}

We now briefly discuss the complexity of the algorithm:
\begin{theorem}\label{thm:complexityFGLM}
The number of operations in $\kk$ of Algorithm~\ref{matricial FGLM} is bounded from above by $O(nD^3)$. 
\end{theorem}
\begin{proof}
We note that all the monomials tested are elements of the staircase $\mathcal{W}_{<_2}\left(\mathcal{G}_2\right)$ or elements of the border $\mcF_{<_2}\left(\mathcal{G}_2\right)$. Since $\left|\mathcal{W}_{<_2}\left(\mathcal{G}_2\right)\right| = D$ and $\left|\mcF_{<_2}\left(\mathcal{G}_2\right)\right| \leq n D$, the algorithm terminates in at most $n D + D$ steps.
All the operations inside the while loop are linear algebra operations such that multiplication between matrices and vectors and elementary operations on the matrix $P$ in the Update procedure. Those operations are in $O(D^2)$. The bound immediately follows from these two observations.
\end{proof}
\subsection{Implementation in CoCoALib}

In this section we show the keypoints of the FGLM algorithm, first in mathematical abstraction and then as the design of the actual implementation.

The mathematical abstraction of the algorithm takes us back in time to the first algorithm using Linear Algebra for computing Gr\"obner bases of zero-dimensional ideals. Buchberger and M\"oller, \cite{BuchbergerMoeller82} showed how to compute, directly from a set of points $\mathbb X\subset \kk^n$, the Gr\"obner basis of the ideal of $\mathbb X$, 
i.e. the zero-dimensional ideal~$I$ of the polynomials vanishing when evaluated in the points in $\mathbb X$.  The abstract structure of this algorithm is the same as in FGLM algorithm, that is:

Consider each power-product $t\in\TT\setminus \LT(I)$, starting from 1 and following the term-ordering:
\begin{itemize}
\item look if there is a linear relation
%    \vskip-0.5cm
    \item if so, this gives a polynomial for the Gr\"obner basis of $I$ with leading term $t$
%    \vskip-0.5cm
    \item if not, $t$ is in the $\kk$-basis of $R/I$ (which is finite, because $I$ is zero-dimensional)
\end{itemize}

Following this approach, several other algorithms have been described for performing computations with zero-dimensional ideals based on linear algebra, see for example 
\cite{MarMoeMor93},
\cite{AbbBigKreRob00}, 
\cite{AbbKreRob05} (on ideals of points, projective points, functionals),
\cite{AbbFasTor08} (for Border bases of ideals of points given with approximate coordinates),
\cite{AbbBigPalRob19} (on the minimal polynomial within a zero-dimensional ideal).

Many of these algorithms have been implemented and highly optimized in CoCoALib (the C++ library which is the mathematical core of the computer algebra system CoCoA-5 \cite{CoCoA}, \cite{CoCoALib}).

So CoCoALib provides two C++ classes developed for this fascinating family of algorithms: \mycode{QBGenerator} and \mycode{LinDepMill}, two \enquote{black-boxes} with a dedicated minimal interface for inserting and extracting useful information.

\medskip
Class \mycode{QBGenerator}:
produces \enquote{the next monomial to consider} in the staircase.

Let $I \subseteq R = \kk[x_1, \dots, x_n]$ be a zero-dimensional ideal.

We seek the \textit{staircase} of $I$ according to the given term-ordering.  This is a $\kk$-basis for the quotient $R/I$, and in this section we call it QB, as in CoCoALib (instead of $S$, as in the previous section).

We want to determine the terms in QB one at a time, starting with the trivial term, 1.  At every stage the partially formed QB is factor-closed (i.e. if $t\in$ QB then any factor of $t$ is in QB), and this implies that, instead of $\TT\setminus QB$, only a few terms are suitable for being considered in the next stage: we call them \textbf{corners}.
We get this list from the QBGenerator by calling \mycode{QBG.myCorners()} and then we can choose one term $t$ (the smallest, for making a Gr\"obner basis).
If we determine that $t$ is in QB (no linear relation) then we call \mycode{QBG.myCornerPPIntoQB(t)}, otherwise (when $t$ is in the leading terms in $I$) we call \mycode{QBG.myCornerPPIntoAvoidSet(pp)}.
These last two functions will update the internal state of the QBGenerator so that at the next stage we will get the new set of corners, excluding all multiples of the terms in the \textit{avoid set}.

\medskip
Class \mycode{LinDepMill}:
seeks a linear dependance in the collected vectors.

A LinDepMill stores an expandable matrix representing \enquote{a running Gaussian elimination}, while we append a row at a time, and it efficiently keeps track of the previous computations.
The matrix itself is intentionally not accessible, because what we just need to know is whether or not there is a linear relation, and which.

By calling \mycode{LDM.myAppendVec(v)} we append a (row) vector to the matrix, then internally it will be Gaussian-reduced.
Then, calling \mycode{LDM.myLinReln()} we will get the coefficients of a linear relation (if any exists) of the last vector with the previous ones.

\textit{For the curious reader, there are two internal implementations of \mycode{LinDepMill}: one is optimized for computations with coefficients in $\FF_p$, the other is general for any ring.   The best one is called automatically.}

For helping the reader interested in understanding the code 
in  CoCoALib,
we write in pseudo-code the structure of our implementation, emphasizing the actual snippets of code
(see file \mycode{SparsePolyOps-ideal-FGLM.C}).
\textit{Please note that in CoCoALib the terms are considered as elements of the \mycode{PPMonoid} $\TT$, instead of ring elements in~$R$.  So sometimes one needs to explicitly turn the terms into polynomials.}

%%% --------------- algorithm ----------------------------------------------
\newcommand{\mycomment}[1]{\textcolor{magenta}{{// #1}}}
\begin{algorithm}[H]\label{CoCoALib FGLM}
		\caption{FGLM algorithm in CoCoALib}
		\begin{algorithmic}
        %% ----- input ----------------------------------
			\Require \\
            $I \subsetneq R_{old}$ an ideal, with $<_{old}$\\
            $R_{new}$ a polynomial ring, with $<_{new}$ a new term order (same coeff ring $K$ and vars as $R_{old}$).
	    %% ----- output ----------------------------------
		\Ensure \\
        {GB}$_{new}$ a reduced Gr\"{o}bner basis of $I$ w.r.t.\ $<_{new}$\\	
        {QB}$_{new}$ a $K$-basis of $R_{new}/I$ w.r.t.\ $<_{new}$			
	
        \State 
        \mycomment{------------ INITIALIZATION -----------------}

        \State QB$_{old}$ = \mycode{QuotientBasisSorted(I)};
        \quad\mycomment{compute QB$_{old}$ from the GBasis w.r.t $<_{old}$} 
            
		\State \mycode{LDM(K, len(QB\_old))};
        \quad\mycomment{create a LinDepMill called LDM in $K^s$}
            
		\State coeffs = coefficients of $t=1$  w.r.t. QB$_{old}$;
        \quad\mycomment{coeffs = (1,0,0,0,0...0) in $K^s$}
            
		\State \mycode{LDM.myAppendVec(coeffs)};
        \quad\mycomment{initialize LDM first row in reducing matrix}

        \State \mycode{QBG(T\_new)}; 
        \quad\mycomment{create a QBGenerator called QBG in the monoid $\TT_{new}$ of $R_{new}$}

        \State \mycode{QBG.myCornerPPIntoQB(one(T\_new))}; 
        \quad\mycomment{initialize QBG with $t=1$}

%QB_new.push_back(one(R_new)); // copy of QB in R_new (polys instead of terms)
        \State \mycomment{------------ MAIN CYCLE -----------------}

        \While {\mycode{QBG.myCorners} is not empty}

        \State $t$ = first term of \mycode{QBG.myCorners}; 
        \quad\mycomment{$t = \min_{<_{new}}$(\texttt{QBG.myCorners})}	

        \State coeffs = coefficients of $NF_{old}(t)$  w.r.t. QB$_{old}$;
        \quad\mycomment{coeffs in $K^s$}

        \State \mycode{LDM.myAppendVec(coeffs)};
        \quad\mycomment{add row to matrix, and reduce matrix}

        \If {\mycode{LDM.myLinReln} is empty} 
        \quad\mycomment{if there is no linear relation}

        \State \mycode{QBG.myCornerPPIntoQB(t)}
        \quad\mycomment{add $t$ to QB$_{new}$, and consequently update corners}

        \Else \quad\mycomment{there is a linear relation}
        \State add the corresponding poly to GB$_{new}$
        \State \mycode{QBG.myCornerPPIntoAvoidSet(t)};
        \quad\mycomment{to avoid all multiples of $t$ from future corners}

        \EndIf

        \EndWhile

		\State\Return GB$_{new}$, QB$_{new}$
	\end{algorithmic}
\end{algorithm}

Thanks to the two underlying classes \mycode{QBGenerator} and \mycode{LinDepMill}, which were already implemented in CoCoALib, the actual code for the function \mycode{FGLM}, is quite straightforward and fits in less than one screen.  
We find this fact quite extraordinary.

%%%%%%%%%%%%%%%%%%%%%%%%%%%%%%%%%%%%%%%%%%%%%%%%%%%%%%%%%%%%%%%%%%%
%%%%%%%%%%%%%%%%%%%%%%%%%  Sezione %%%%%%%%%%%%%%%%%%%%%%	
\newpage
\section{F4 Algorithm}\label{sec F4}
		
In this section we present the F4 Algorithm, a linear algebra based algorithm for computing Gr\"obner basis, introduced by Faugère in \cite{F4paper}.

\noindent For this section we fix a term order $\tau$ on the polynomial ring $R=\kk[x_1,\dots,x_n]$.

\subsection{Preliminaries}

\begin{definition}\label{defnewvar}Let $A \in M_{\ell,s}(\kk)$ and $t = [y_1 >_{\Lex} \cdots >_{\Lex} y_s]$ be new variables. Then $F = \mbox{rows}_t(A)$ is a set of linear forms; and we can compute its reduced Gr\"obner basis $\Tilde{F}$ with respect to the term order $\Lex$ on $y_1 >_{\Lex} \cdots >_{\Lex} y_s$. From this basis $\Tilde{F}$ we can construct $\mbox{matrix}(\Tilde{F}).$ We notice that $\mbox{matrix}(\Tilde{F})$ is the RREF of the matrix $A$. We say that $\Tilde{F}$ is a \textbf{row echelon basis} of $F$, or also that $\Tilde{F}$ is the \textbf{RREF} (\textbf{reduced row echelon form}) of $F$:
$$\Tilde{F} = \rows_{[y_1 >_{\Lex} \cdots >_{\Lex} y_s]}(\mbox{RREF}(A)).$$
\end{definition}

In the case of polynomials we have a similar definition.

\begin{definition}
Let $F \subseteq R$ be a finite list of elements of $R$. We take the list of terms $t = \Supp(F)$ ordered in a descending manner according to $\tau$. Let $\Tilde{A}$ be the RREF of $A = \mbox{matrix}(F).$ We say that $\Tilde{F} = \mbox{rows}_t(\Tilde{A})$ is the \textbf{reduced row echelon form} (\textbf{RREF}) of $F$ with respect to the term order $\tau$:
$$\Tilde{F} = \mbox{rows}_{\Supp(F)}(\mbox{RREF}(\mbox{matrix}(F))).$$
\end{definition}

\begin{remark}
From the previous definition, it follows immediately that $\Supp(\Tilde{F}) = \Supp(F)$.
\end{remark}

\begin{example}
If we take $F = [x^5 y + 5x^2 y + 94 x, x^5 y + 100 y] \subseteq \mathbb{F}_{101}[x,y]$, we have that $\Supp(F) = [x^5 y >_{\tau} x^2 y >_{\tau} x >_{\tau} y].$ So
$$A = \bordermatrix{
    &x^5y & x^2y &x & y\cr
    &1 & 5 &  94 & 0 \cr
    &1 & 0 & 0 & 100} \ \mbox{and}\ \Tilde{A} = \bordermatrix{
    &x^5y & x^2y &x & y\cr
    &1 & 0 & 0 & 100 \cr
    &0 & 1 & 39 & 81};$$
then $\Tilde{F} = \{x^5 y + 100 x, x^2 y + 39 x + 81 y\}.$
\end{example}

Elementary properties of row echelon matrices are summarized by the following theorem.

\begin{theorem}\label{thmtriangularbasis}
Let $A \in M_{\ell,s}(\kk)$ and $t = [y_1 > \cdots > y_s]$ be new variables with the term order $\Lex$; let $F = \rows_t(A)$, $\Tilde{A}$ the RREF of $A$, and $\Tilde{F} = \rows_t(\Tilde{A}).$ We define
$$\Tilde{F}^+ := \{g \in \Tilde{F} \ | \ \LT_{\Lex}(g) \not\in \LT_{\Lex}(F) \}.$$
For any subset $F_-$ of $F$ such that $|F_-| = |\LT_{\Lex}(F)|$ and $\LT_{\Lex}(F_-) = \LT_{\Lex}(F)$, then $G := \Tilde{F}^+ \cup F_-$ is a triangular basis of the $\kk$-vector space $\langle F \rangle_{\kk} = \rowsp_t(A)$. That is to say, for all $f \in \langle F \rangle_{\kk}$ there exist $(\lambda_k)_{k = 1,\dots,s}$ elements of $\kk$ and $(g_k)_{k = 1,\dots,s}$ elements of $G$ such that
$$f = \sum\limits_{k=1}^s \lambda_k g_k, \ \LT_{\Lex}(g_1) = \LT_{\Lex}(f), \ \mbox{and $\LT_{\Lex}(g_k) >_{\Lex} \LT_{\Lex}(g_{k+1})$ for $k =1,\dots,s-1$}.$$
\end{theorem}
\begin{proof}
By definition,  $F$ is a system of generators for $\langle F \rangle_{\kk}$. Because of the properties of row echelonization, we have that $\Tilde{F}$ is a basis for $\langle F \rangle_{\kk}$; $\Tilde{F}$ is also a triangular, because the leading terms of the polynomials $\Tilde{F}$ are pairwise distinct.\\
Now, for the latter argument, $\Tilde{F}^+$ is a basis for $\langle \Tilde{F}^+ \rangle_{\kk}$ and $F_-$ is a basis for the $\kk$-vector subspace $C$ such that $C \subseteq \langle F \rangle_{\kk}$ and $C \oplus \langle \Tilde{F}^+ \rangle_{\kk} = \langle F \rangle_{\kk}$. So $G = \Tilde{F}^+ \cup F_-$ is a system of generators for
$$\langle \Tilde{F}^+ \rangle_{\kk} + C = \langle \Tilde{F}^+ \rangle_{\kk} \oplus C = \langle F \rangle_{\kk}.$$
Finally, because the leading terms of the elements in $G$ are pairwise distinct, $G$ is a basis of $\langle F \rangle_{\kk}$ that is also triangular.
\end{proof}

We can immediately translate the previous theorem into a result for polynomials.

\begin{corollary}\label{cortriangularbasis}
Let $F$ be a finite subset of $R$ and $\tau$ a term order. Let $\Tilde{F}$ be the RREF of $F$ with respect to $\tau$. We define 
$$\Tilde{F}^+ := \{g \in \Tilde{F} \ | \ \LT_{\tau}(g) \not\in \LT_{\tau}(F) \}.$$
For all subset $F_-$ of $F$ such that $|F_-| = |\LT_{\tau}(F)|$ and $\LT_{\tau}(F_-) = \LT_{\tau}(F)$, then $G := \Tilde{F}^+ \cup F_-$ is a triangular basis of the $\kk$-vector space $\langle F \rangle_{\kk}$.
\end{corollary}
\begin{proof}
We choose $t = \Supp(F) = [t_1 >_{\tau} \cdots >_{\tau} t_s]$. Then, we have the following isomorphism of $\kk$-vector spaces
\begin{align*}
    \psi \colon \ \langle y_1 >_{\Lex} \cdots >_{\Lex} y_s &\rangle_{\kk} \xrightarrow{} \langle t_1 >_{\tau} \cdots >_{\tau} t_s \rangle_{\kk}\\
    &y_i \longmapsto t_i, 
\end{align*}
which also preserves the order of the terms. So, if we combine this change of basis with Theorem~\ref{thmtriangularbasis}, we get the desired result.
\end{proof}

\begin{example}
We consider $R=\mathbb{F}_{101}[x,y,z]$ with $\tau$ the $\DegRevLex$ term order and the list
\[
F = [f_1 = x^3 + 11 xy^2 + 17 xyz + 12 x + 87z, f_2 = x^3 +22 y^2z + 35y, f_3 = x^2 + 93 z, f_4 = x^2 + 12 x + y].
\]  Then, we have
$$\asmatrix(F) = \bordermatrix{
    &x^3 & xy^2 & xyz & y^2z & x^2 & x & y & z\cr
    f_1 &1 & 11 & 17 & 0 & 0 & 12 & 0 & 87\cr
    f_2 & 1 & 0 & 0 & 22 & 0 & 0 & 35 & 0\cr
    f_3 & 0 & 0 & 0 & 0 & 1 & 0 & 0 & 93\cr
    f_4 & 0 & 0 & 0 & 0 & 1 & 12 & 1 & 0}.$$
So $\Tilde{F} = \rows_{\Supp(F)}(\mbox{RREF}(\asmatrix(F))).$ We have that
$$\mbox{RREF}(\asmatrix(F)) = \bordermatrix{
    &x^3 & xy^2 & xyz & y^2z & x^2 & x & y & z\cr
    \Tilde{f_1} &1 & 0 & 0 & 22 & 0 & 0 & 35 & 0 \cr
    \textcolor{blue}{\Tilde{f_2}} & 0 & \textcolor{blue}{1} & 75 & 99 & 0 & 0 & 61 & 99 \cr
    \Tilde{f_3} & 0 & 0 & 0 & 0 & 1 & 0 & 0 & 93 \cr
    \textcolor{blue}{\Tilde{f_4}} & 0 & 0 & 0 & 0 & 0 & \textcolor{blue}{1} & 59 & 68}.$$
\end{example}
So $\Tilde{F} = [\Tilde{f_1} = x^3 + 22 y^2 z + 35 y, \Tilde{f_2} = xy^2 + 75 xyz + 99 y^2z + 61 y + 99z, \Tilde{f_3} = x^2 + 93 z, \Tilde{f_4} = x + 59 y + 68 z].$
Now, the \enquote{new} leading terms of $\Tilde{F}$ are the ones marked in blue above, so 
$$\Tilde{F}^+ = \{\Tilde{f_2} = xy^2 + 75 xyz + 99 y^2z + 61 y + 99z, \Tilde{f_4} = x + 59 y + 68 z\}.$$
Then, if we take $F_- = \{f_1 = x^3 + 11 xy^2 + 17 xyz + 12 x + 87z, f_3 = x^2 + 93 z\}$, all assumptions of Theorem~\ref{thmtriangularbasis} are satisfied, and $G = \Tilde{F}^+ \cup F_- = \{f_1,\Tilde{f_2},f_3,\Tilde{f_4}\}$ is the triangular basis that we want for $\langle F \rangle_{\kk}:$
$$\asmatrix(G) = \bordermatrix{
    &x^3 & xy^2 & xyz & y^2z & x^2 & x & y & z\cr
    f_1 &\textcolor{red}{1} & 11 & 17 & 0 & 0 & 12 & 0 & 87\cr
    \Tilde{f_2} & 0 &  \textcolor{red}{1} & 75 & 99 & 0 & 0 & 61 & 99 \cr
    f_3 & 0 & 0 & 0 & 0 & \textcolor{red}{1} & 0 & 0 & 93\cr
    \Tilde{f_4} & 0 & 0 & 0 & 0 & 0 & \textcolor{red}{1} & 59 & 68}.$$
Observe that we could have also chosen $F_- =\{f_2,f_4\}$, $F_- = \{f_1,f_4\}$, or $F_- = \{f_2,f_3\}$.

\subsection{F4}

We are now ready to describe the $\Ffour$ Algorithm. It is well known that during the execution of Buchberger's Algorithm, several choices must be made—specifically, the selection of a critical pair from the list of critical pairs, and the choice of a reductor from a list of candidates when reducing a polynomial by a set of polynomials.
While these choices do not affect the correctness of the algorithm, they are crucial for its overall computational efficiency. In practice, the most effective strategies rely solely on the leading terms of the polynomials to guide these decisions. However, when all input polynomials share the same leading term, all critical pairs become indistinguishable, making it impossible to apply such heuristics.
This issue can be addressed in a simple yet effective way: by making no choice at all. That is, instead of selecting a single critical pair at each step, a subset of critical pairs is processed simultaneously. In this way, the burden of decision-making is deferred to a later stage of the algorithm—the linear algebra phase.

\begin{definition}
A \textbf{critical pair} of two polynomials $(f_i,f_j) \in R^2$ is an element of $\TT \times R^2$, $\Pair_{\tau}(f_i,f_j) := (\lcm_{ij},\shalf(f_i,f_j),\shalf(f_j,f_i))$, such that
\[
\lcm(\Pair_{\tau}(f_i,f_j)) = \lcm_{i,j} = \LT_{\tau}(\shalf(f_i,f_j)) = \LT_{\tau}(\shalf(f_j,f_i)) = \lcm(\LT_{\tau}(f_i),\LT_{\tau}(f_j)).
\]

The \textbf{degree of the critical pair} is $\deg(p_{i,j}):=\deg(\lcm_{i,j})$, for $p_{i,j}=\Pair_{\tau}(f_i,f_j)$.
\end{definition}

The $\Ffour$ Algorithm as explained in \cite{F4paper} consists indeed of three algorithms: one called $\Ffour$, one called Reduction ($\Red$) and one called Symbolic Preprocessing ($\symPre$).

\begin{algorithm}[H]
\caption{$\Ffour$} 
\label{F4 alg}
\begin{align*}
    &\mbox{\textbf{Input}}: \begin{cases}
        \mbox{$F$ a finite list of elements of $R$;} \\
        \mbox{$\Sel: \List(\Pairs) \to \List(\Pairs)$ a function such that $\Sel(\ell) \not= \emptyset$ if $\ell\not= \emptyset$;} \\
        \mbox{a term order $\tau$.}
    \end{cases}\\
    &\mbox{\textbf{Output}: a finite list of elements of $R$ that is a Gr\"obner basis for $(F)$ w.r.t.\ $\tau$.}\\
    & G := F, \  \Tilde{F}_0^+ := F, \ d := 0; \\
    & P := [\Pair_{\tau}(f,g) \ | \ (f,g) \in G^2 \ \mbox{with} \ f\not=g]; \ \  \mbox{\textcolor{magenta}{//if we consider $(f,g)$ we discard $(g,f)$}}  \\
    &\mbox{\textbf{while} $P \not= \emptyset$ \textbf{do}}\\
    & \ d := d+1;\\
    & \ P_d := \Sel(P); \\
    & \ P := P \smallsetminus P_d;\\
    & \ L_d := [\shalf(f_i,f_j),\shalf(f_j,f_i) \ |\ p_{i,j} = \Pair_{\tau}(f_i,f_j) \in P_d];\\
    & \ \Tilde{F}_d^+ := \Red(L_d,G);\\
    & \ \mbox{\textbf{for} all $h \in \Tilde{F}_d^+$ \textbf{do}}\\
    & \ \ P := P \cup [\Pair_{\tau}(h,g) \ | \ g \in G];\\
    & \ \ G := G \cup [h];\\
    & \ \mbox{\textbf{end for};}\\
    & \mbox{\textbf{end while};}\\
    & \mbox{\textbf{return} $G$}
\end{align*}
\end{algorithm}

Now, we describe the other two sub-routines that are needed in the execution of  the algorithm: $\Red$ and $\symPre$.
The $\Red$ Algorithm describes how to reduce a set of polynomials using another set of polynomials. This corresponds to the linear algebra phase of the algorithm, where the S-polynomials from Buchberger's Algorithm are reduced according to a quantity determined by the function $\Sel$.

\begin{algorithm}[H]
\caption{$\Red$}
\label{Reduction}
\begin{align*}
    &\mbox{\textbf{Input}: $\begin{cases}
            \mbox{$L,G$ finite lists of elements of $R$;}\\
        \mbox{a term order $\tau$.}
    \end{cases}$}\\
    &\mbox{\textbf{Output}: a finite list of elements of $R$ (eventually empty).}\\
    & F := \symPre(L,G);\\
    & \mbox{$\Tilde{F} :=$ Reduction to RREF of $F$ with respect to $\tau$} ;\\
    & \Tilde{F}^+ := [f \in \Tilde{F} \ | \ \LT_{\tau}(f) \not\in \LT_{\tau}(F)]; \ \textcolor{magenta}{// \ \mbox{``useful" piece of $F$}}  \\
    & \mbox{\textbf{return} $\Tilde{F}^+$}
\end{align*}
\end{algorithm}

The final subroutine of the $\Ffour$ algorithm is also its most important, as it constructs the set of polynomials used to reduce the selected critical pairs. In other words, it is responsible for generating the matrix, for which we will compute the reduced row echelon form (RREF) with respect to the chosen term order $\tau$. Since this step involves no arithmetic operations, it can be seen as a symbolic preprocessing phase, hence the name $\symPre$.

\begin{algorithm}[H]
\caption{$\symPre$}
\label{symPre}
\begin{align*}
    &\mbox{\textbf{Input}: $\begin{cases}
            \mbox{$L,G$ finite lists of elements of $R$;}\\
        \mbox{a term order $\tau$.}
    \end{cases}$}\\
    &\mbox{\textbf{Output}: a finite list of elements of $R$.}\\
    & F := L;\\
    & \Done := \LT_{\tau}(F) ;\\
    &\mbox{\textbf{while} $\Supp(F) \not= \Done$ \textbf{do}}\\
    & \ \mbox{Select $t$ an element of $\Supp(F) \smallsetminus \Done$}; \\
    & \ \Done := \Done \cup [t]; \\
    & \ \mbox{\textbf{if} there exist $g \in G$ and $t' \in \TT$ such that $t = t' \cdot \LT_{\tau}(g)$ \textbf{then}}\\
    & \ \ F := F \cup [t' \cdot g];\\
    & \ \mbox{\textbf{end if};}\\
    & \mbox{\textbf{end while};}\\
    & \mbox{\textbf{return} $F$}
\end{align*}
\end{algorithm}

\begin{remark}
We show now that the procedure $\symPre$ ends.
At the beginning of the algorithm, 
we build a rooted tree as follows. We denote with $\bullet$ the root and let its children be $\Supp(F) \setminus \Done = \left\{t_1, \dots, t_n\right\}$.
We  dispose all the terms that appear at a certain step of the algorithm in $\Supp(F) \setminus \Done$, in  $n$ sub-trees whose roots are precisely $t_1, \dots, t_n$ and  we want to show that the whole tree must be finite. Suppose indeed that at the first step we select $t_1$ and find $g \in G$ such that $t_1 = t' \LT(g)$ for some $t' \in \TT$ and add $t'g$ to $F$. The new terms of $\Supp(F) \setminus \Done$ are of the form $t'a$ where $a \in \Supp(g) \setminus \left\{\LT(g)\right\}$. Since $t_1 = t'\LT(g)$, all the terms that we are adding are strictly lower than $t_1$ with respect to the term order $\tau$. We put the new terms say $t'a_1, \dots, t'a_{k_1}$ as children of $t_1$ (i.e as $k_1$ leaves $ta_1, \dots, ta_{k_1}$). At each step, for a tested term $t$ we add the new terms that has been generated as children of $t$. Proceeding in this way, we are building a tree where all the nodes have a finite number of children.\\
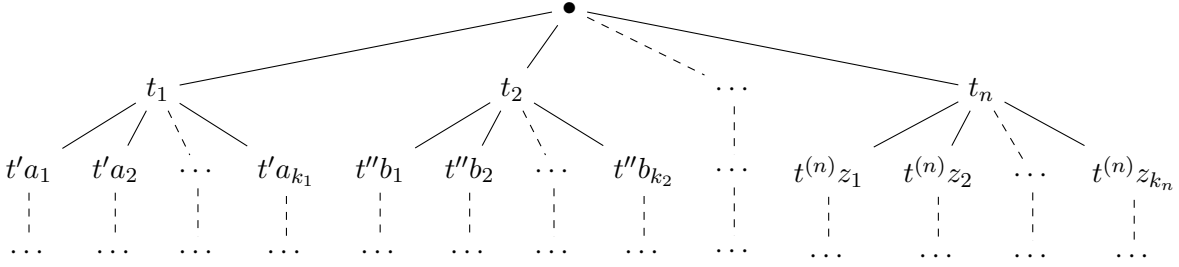
\begin{center}
\begin{forest}
  [$\bullet$
    [$t_1$
      [$t'a_1$
      [$\cdots$, edge = dashed]]
      [$t'a_2$
      [$\cdots$, edge = dashed]]
      [$\cdots$, edge = dashed
      [$\cdots$, edge = dashed]]
      [$t'a_{k_1}$
      [$\cdots$, edge = dashed]]
    ]
    [$t_2$
      [$t''b_1$
      [$\cdots$, edge = dashed]]
      [$t''b_2$
      [$\cdots$, edge = dashed]]
      [$\cdots$, edge = dashed
      [$\cdots$, edge = dashed]]
      [$t''b_{k_2}$
      [$\cdots$, edge = dashed]]
    ]
    [$\cdots$,edge = dashed
      [$\cdots$, edge = dashed
      [$\cdots$, edge = dashed]]
    ]
    [$t_n$
      [$t^{(n)}z_1$
      [$\cdots$, edge = dashed]]
      [$t^{(n)}z_2$
      [$\cdots$, edge = dashed]]
      [$\cdots$, edge = dashed
      [$\cdots$, edge = dashed]]
      [$t^{(n)}z_{k_n}$
      [$\cdots$, edge = dashed]]
    ]
  ]
\end{forest}
    
\end{center}

Suppose by contradiction that the \textbf{while} loop of the $\symPre$ algorithm does not end. Then, the tree we are building is infinite, because we are indefinitely adding a finite quantity of leaves to our tree. Then, thanks to K{\H{o}}nig's Lemma (see \cite{K01}), our tree must have an infinite descending path, i.e.\ an infnite strictly-descending sequence of terms: this contradicts Lemma \ref{descending chains}. $\lightning$
\end{remark}

\subsection{Termination of the F4 Algorithm}

Now, we explain why the $\Ffour$ algorithm terminates and produces a Gr\"obner basis for the desired ideal.
We begin with a couple of lemmas.

\begin{lemma}\label{lemF4termination}
Let $G \subseteq R$ be a finite list of polynomials, let $\tau$ be a term order, let $s \geq 1$ an integer and let
$$L := [t_1 \cdot g_1, \cdots, t_s \cdot g_s \ | \ t_1,\dots,t_s \in \TT, g_1,\dots,g_s \in G],$$
where the $t_i$'s and $g_i$'s are not forced to be pairwise distinct (so it can be that $s > |G|$). Moreover, let $\Tilde{F}^+ := \Red(L,G),$ and let $J = (G)$ the ideal generated by $G$. Then, for all $h \in \Tilde{F}^+$,
$$\LT_{\tau}(h) \not\in \LT_{\tau}(J).$$
\end{lemma}

\begin{proof}
Let $F := \symPre(L,G)$. Suppose by contradiction that there exists $h \in \Tilde{F}^+$ such that $\LT_{\tau}(h) \in \LT_{\tau}(J)$. Then there exists $g \in G$ such that $\LT_{\tau}(g) | \LT_{\tau}(h)$, and, moreover, $\LT_{\tau}(h) \not\in \LT_{\tau}(F)$.\\
Then, we have that
$$\LT_{\tau}(h) \in \Supp(\Tilde{F}^+) \subseteq \Supp(\Tilde{F}) = \Supp(F),$$
and $\LT_{\tau}(g) | \LT_{\tau}(h)$; so $\symPre$ puts in $F$ the product $\frac{\LT_{\tau}(h)}{\LT_{\tau}(g)} \cdot g$, or a product with the same head term (that is to say with head term equal to $\LT_{\tau}(h)$). Then, $\LT_{\tau}(h) \in \LT_{\tau}(F)$, and this is a contradiction. $\lightning$
\end{proof}

\begin{lemma}\label{lemF4correctness}
Let $G, \tau$, and $L$ be as in Lemma~\ref{lemF4termination}.
Let $\Tilde{F}^+ := \Red(L,G).$ Then $\Tilde{F}^+$ is a subset of $J = (G)$ (the ideal generated by $G$). Moreover, for all $f$ in the $\kk$-vector space gerated by $L$ we have that the total reduction of $f$ via $G \cup \Tilde{F}^+$ is $0$ (that is to say,
$\multiVarDiv(f, G \cup \Tilde{F}^+) = 0$).
\end{lemma}
\begin{proof}
Apply the Corollary \ref{cortriangularbasis} to the set $F$ generated by $\symPre(L,G)$. Clearly, $F$ is a subset of $F \cup J$, but it is obvious that $L$ is a subset of $J$, so that $F$ is a subset of $J$. Hence any $F_-$ fulfilling the hypothesis of Corollary \ref{cortriangularbasis} is a subset of $J$. Then $\Tilde{F}^+$ is a subset of $J$. This concludes the proof because the $\kk$-vector space generated by $L$ is a sub-vector space of the $\kk$-vector space generated by $F$ (and every element in the $\kk$-vector space generated by $F$ has total reduction via $G \cup \Tilde{F}^+$ equal to $0$). 
\end{proof}

\begin{remark}
Let $G$ be a finite subset of $R$. It is possible that for an $f \in R$ the total reduction via $G$ is $0$ (i.e.\ $\multiVarDiv(f,G) = 0$) but the Buchberger reduction via $G$ is non-zero. The reason is that the Buchberger reduction algorithm depends on many choices and strategies.
\end{remark}

We are now ready to prove the correctness and termination of the $\Ffour$ Algorithm.

\begin{theorem}
Let $\tau$ be a term order. The algorithm $\Ffour$ computes a Gr\"obner basis $G \subseteq R$ of an ideal $(F) \subseteq R$ w.r.t.\ $\tau$, such that $F \subseteq G$ and $(F) = (G)$. 
\end{theorem}
\begin{proof}
We begin by investigating the termination of the $\Ffour$ Algorithm. Assume by contradiction that the \textbf{while} loop does not terminate. Then, there exists an ascending sequence $(d_i)$ of natural numbers such that $\Tilde{F}^+_{d_i} \not= \emptyset$ for all $i$. Let $q_i \in \Tilde{F}^+_{d_i}$ for all $i$ (hence $q_i$ can be any element of $\Tilde{F}^+_{d_i}$). Let $U_i := U_{i-1} + (\LT_{\tau}(q_i))$ for $i \geq 1$ and $U_0 := (0)$. By Lemma \ref{lemF4termination} we have that $U_{i-1} \subsetneq U_i$ for all $i \geq 1$. This contradicts the fact that $R$ is Noetherian. $\lightning$\\
We now investigate the correctness of $\Ffour$. At the end of the algorithm, we have that $G = \bigcup\limits_{d \geq 0} \Tilde{F}^+_{d}$. We claim that the following are loop invariants of the \textbf{while} loop: $G$ is a finite subset of $R$ such that  $F \subseteq G \subseteq (F)$, and $\multiVarDiv(\sPol_{\tau}(g_1,g_2),G) = 0$ for all $(g_1,g_2) \in G^2$ such that $(g_1,g_2) \not\in P$. Remember that 
$$\sPol_{\tau}(g_1,g_2) :=  \frac{\LM_{\tau}(g_2)}{\gcd(\LT_{\tau}(g_1),\LT_{\tau}(g_2))} g_1 - \frac{\LM_{\tau}(g_1)}{\gcd(\LT_{\tau}(g_1),\LT_{\tau}(g_2))}g_2.$$
The first claim is an immediate consequence of the Lemma \ref{lemF4correctness}. For the second one, if $\Pair_{\tau}(g_1,g_2) \not\in P$, this means that $\Pair_{\tau}(g_1,g_2) = (\lcm_{1,2},\shalf(g_1,g_2),\shalf(g_2,g_1))$ has been selected in a previous step (say $d$) by the function $\Sel$. Hence $\shalf(g_1,g_2)$ and $\shalf(g_2,g_1)$ are in $L_d$, so $\sPol_{\tau}(g_1,g_2)$ is an element of the $\kk$-vector space generated by $L_d.$ Hence, by the second part of Lemma \ref{lemF4correctness}, the total reduction of $\sPol_{\tau}(g_1,g_2)$ via $G$ is equal to $0$. \\
Remember that $G \subseteq R$ is a Gr\"obner basis (for $(G)$) if and only if
$$\multiVarDiv(\sPol_{\tau}(g_1,g_2),G) = 0$$
for all $g_1,g_2 \in G, g_1 \not= g_2$. But, because the $\Ffour$ Algorithm terminates, the \textbf{while} loop will eventually end, and so, eventually, $P = \emptyset$. So, in the end, $F \subseteq G \subseteq (F)$ and 
$$\multiVarDiv(\sPol_{\tau}(g_1,g_2),G) = 0$$
for all $(g_1,g_2) \in G^2,$ i.e.\ $G$ is a Gr\"obner basis for $(F)$ w.r.t.\ $\tau$.
\end{proof}

\begin{remark}
The $\Ffour$ algorithm, as presented above in its core form, is somewhat slow. This is mainly due to the fact that we keep the generators of the ideal in the Gr\"obner basis, and when we add new polynomials to the working Gr\"obner basis we do not do any reduction among them. We miss the interreduction $\interred$ (Algorithm~\ref{interredltfirst}). 
Therefore, in our implementation in Sage we apply $\interred$ at the beginning and at the end, so  we start with interreduced polynomials and we end with interreduced ones, that is a reduced Gr\"obner basis. During the execution of the \textbf{while} loop, when we add new polynomials, we also perform some mutual reductions via $\multiVarDiv$. However,  we do not do a complete interreduction, since otherwise we could not keep track of the pairs that we have not processed yet and we would have to recompute all pairs altogether. 
\end{remark}

\subsection{Selection Strategies}\label{sec:strategies}
One of the inputs to Algorithm~\ref{F4 alg} is a selection function  $\Sel: \List(\Pairs) \to \List(\Pairs)$ which is used to select the pairs to be considered. This introduces a degree of flexibility into the algorithm, as different selection functions $\Sel$ can lead to significantly different behaviors of the $\Ffour$ Algorithm.

Two very natural choices for $\Sel$ are the following.
\begin{enumerate}
\item  If $\Sel$ is the identity function, then we really reduce all the critical pairs at the same time in each cycle of the \textbf{while} loop. This is usually not the best option because it leads to very large non-sparse matrices which one has to compute Gaussian elimination on.
\item If $|\Sel(\ell)|=1$ for all $\ell \not= \emptyset$, then the $\Ffour$ Algorithm is just the Buchberger Algorithm (Algorithm~\ref{Buchberger}). In this case, the $\Sel$ function corresponds to the selection strategy.
\end{enumerate}

In \cite{F4paper}, Faugère suggests a selection function, called \textbf{normal strategy} for $\Ffour$, which takes all the critical pairs with a minimal total
degree, i.e.\ taking $\Sel$ as presented in Algorithm~\ref{alg:normalstrategy}.
\begin{algorithm}
\caption{Normal Strategy}\label{alg:normalstrategy}
\begin{align*}
    & \mbox{$\Sel(P)$ as selection of critical pairs with minimal total degree} \\
    &\mbox{\textbf{Input}: $P$ a list of critical pairs.}\\
    &\mbox{\textbf{Output}: a list of critical pairs.}\\
    & d := \min\{\deg(\lcm(p))  \mid  p \in P\};\\
    & P_d := [p \in P  \mid  \deg(\lcm(p)) = d];\\
    & \mbox{\textbf{return} $P_d$.}
\end{align*}
\end{algorithm}
Notice that, if the input polynomials are homogeneous, we already have a Gr\"obner basis up to degree $d - 1$ and $\Sel$ selects
exactly all the critical pairs which are needed for computing a Gr\"obner basis up to degree $d$.

In Example~\ref{ex:comparisonstragies}, we illustrate the different pairs that result from using different selection strategies. In order to focus on the selection strategies and better observe the differences we do not apply $\interred$ in this example.
\begin{example}\label{ex:comparisonstragies}
We compute the Gr\"obner basis of the ideal
$$(xyz + 100, xz + y^2z) \subseteq \mathbb{F}_{101}[x,y,z]$$
with respect to the term order $\Lex, x >_{\Lex} y >_{\Lex} z$, with three different selection strategies.
\begin{enumerate}
    \item $\Sel = \mbox{Id} : \mbox{List(Pairs)} \longrightarrow \mbox{List(Pairs)}$ the identity function. \\
    The \textbf{while} loop is executed 3 times, and the selected pairs are:
    \begin{align*}
P_1 &= [ \ [xyz, xyz + 100, xyz + y^3z] \ ] \\
P_2 &= [ \ [xy^3z, xy^3z + x, xy^3z + 100y^2], \\
&\ \ \ \ \ \ [xy^3z, xy^3z + x, xy^3z + y^5z] \ ] \\
P_3 &= [ \ [xyz, xyz + y^3z, xyz + 100], \\
&\ \ \ \ \ \ [xz, xz + y^2z, xz + y^2z], \\
&\ \ \ \ \ \ [xy^3z, xy^3z + y^5z, xy^3z + x] \ ].
\end{align*}
    \item $\Sel: \mbox{List(Pairs)} \longrightarrow \mbox{List(Pairs)}, P \longmapsto [\mbox{First}(P)]$ the Buchberger-like function that selects the first pair in all non-empty lists of pairs; \\
    The \textbf{while} loop is executed 6 times, and the selected pairs are:
    \begin{align*}
P_1 &= [ \ [xyz, xyz + 100, xyz + y^3z] \ ] \\
P_2 &= [ \ [xy^3z, xy^3z + x, xy^3z + 100y^2] \ ] \\
P_3 &= [ \ [xy^3z, xy^3z + x, xy^3z + y^5z] \ ] \\
P_4 &= [ \ [xyz, xyz + y^3z, xyz + 100] \ ] \\
P_5 &= [ \ [xz, xz + y^2z, xz + y^2z] \ ] \\
P_6 &= [ \ [xy^3z, xy^3z + y^5z, xy^3z + x] \ ].
\end{align*}
    \item $\Sel: \mbox{List(Pairs)} \longrightarrow \mbox{List(Pairs)}, P \longmapsto \biggl[\underset{p \in P}{\arg\min}\{\deg(\lcm(p))\}\biggr]$ the normal selection strategy. \\
    The \textbf{while} loop is executed 5 times, and the selected pairs are:
    \begin{align*}
P_1 &= [ \ [xyz, xyz + 100, xyz + y^3z] \ ] \\
P_2 &= [ \ [xy^3z, xy^3z + x, xy^3z + 100y^2], \\
&\ \ \ \ \ \ [xy^3z, xy^3z + x, xy^3z + y^5z] \ ] \\
P_3 &= [ \ [xy^3z, xy^3z + y^5z, xy^3z + x] \ ] \\
P_4 &= [ \ [xyz, xyz + y^3z, xyz + 100] \ ] \\
P_5 &= [ \ [xz, xz + y^2z, xz + y^2z] \ ].
\end{align*}
\end{enumerate}
In the three cases the $\Ffour$ Algorithm return the same Gr\"obner basis, which is $\{xyz + 100, xz + y^2z, y^3z + 1, x + y^2\}$. Notice that this basis is not interreduced, 
It is easy to observe that, even in this small example, different selection strategies can lead to the selection of different pairs at each step, particularly in the middle of the algorithm.
\end{example}

%%%%%%%%%%%%%%%%%%%%%%%%%%%%%%%%%%%%%%%%%%%%%%%%%%%%%%%%%%%%%%%%%%%
%%%%%%%%%%%%%%%%%%%%%%%%%  Sezione %%%%%%%%%%%%%%%%%%%%%%	
\newpage
\addtocontents{toc}{\protect\setcounter{tocdepth}{2}}
\section{A Sage Implementation of F4}\label{sec:implementation}
In this section, we give a brief description of our Sage implementation of the F4 algorithm which is available on GitHub at \url{https://github.com/tor-kristian/f4_genova}.

\subsection{Download and Set-up}

\subsubsection*{Requirements}

\begin{itemize}
    \item Sage at \url{https://www.sagemath.org/} 
    \item Python at \url{https://www.python.org/}
    \item Jupyter Notebook at \url{https://jupyter.org/}
\end{itemize}

\subsubsection*{Installation Procedure}

Download the f4.py from GitHub at \url{https://github.com/tor-kristian/f4_genova} and rename the file as f4.ipynb, in order to work on it with Jupyter Notebook \cite{Jupyter} (with the SageMath kernel).

Choose a field and a number of variables. When defining the polynomial ring, choose a term order using the standard Sage command \textcolor[HTML]{006EB8}{PolynomialRing}. Next, insert the polynomial system as a list of polynomials $F$. To compute the reduced Gr\"obner basis of the system, run the algorithm \textcolor[HTML]{006EB8}{F4($F$,ver)}. Notice that the selection strategy implemented is the normal strategy (see Section~\ref{sec:strategies}).

\subsection{Functions}

\textcolor[HTML]{009B55}{\textbf{def}} \textcolor[HTML]{006EB8}{multiVarDiv(f, G):} \newline
We have implemented the algorithm for multivariate division using Algorithm~\ref{PolynomialTotalReduction}. It takes an input polynomial $f$, and reduces it by $G$. The algorithm returns a list of the quotiens polynomials $qP$ and the remainder $r$.
\\ \\ 
\textcolor[HTML]{009B55}{\textbf{def}} \textcolor[HTML]{006EB8}{AscOrdLT(F):} \newline
Reordering algorithm. Returns the reordering of the list of polynomials $F$, from the smallest to the biggest according to the ordering given on the the leading terms of the polynomials by the working term order.
\\ \\ 
\textcolor[HTML]{009B55}{\textbf{def}} \textcolor[HTML]{006EB8}{interred(F):} \newline
Interreduction algorithm (Algorithm~\ref{interredltfirst}). Returns the interreduction of the list of polynomials F, according to the term order we are using.
\\ \\
\textcolor[HTML]{009B55}{\textbf{def}} \textcolor[HTML]{006EB8}{sPol(f,g):} \newline
Computes the S-polynomial of two polynomials $f$ and $g$ as defined in Theorem~\ref{BuchbergerCrit}. This algorithm is not used in our F4 implementation, where we rather use the function  \textcolor[HTML]{006EB8}{ s\_half(f,g)} below.
\\ \\
\textcolor[HTML]{009B55}{\textbf{def}} \textcolor[HTML]{006EB8}{ s\_half(f,g):} \newline
Computes the s-half of two polynomials $f$ and $g$. 
\\ \\
\textcolor[HTML]{009B55}{\textbf{def}} \textcolor[HTML]{006EB8}{ leadingTerms(F):} \newline
Given a set of polynomials $F$, return a list containing all the leading terms of the polynomials in $F$.
\\ \\
\textcolor[HTML]{009B55}{\textbf{def}} \textcolor[HTML]{006EB8}{ leadingMonomials(F):} \newline
Given a set of polynomials $F$, return a list containing all the leading monomials of the polynomials in $F$.
\\ \\
\textcolor[HTML]{009B55}{\textbf{def}} \textcolor[HTML]{006EB8}{getMonomials(f):}\newline
Given a polynomial $f$, return a list containing all monomials of $f$. 
\\ \\
\textcolor[HTML]{009B55}{\textbf{def}} \textcolor[HTML]{006EB8}{getTerms(f):}\newline
Given a polynomial $f$, return a list containing all terms of $f$.
\\ \\
\textcolor[HTML]{009B55}{\textbf{def}} \textcolor[HTML]{006EB8}{maCols(sHalf):}\newline
Given a list of s-halves $sHalf$, return an ordered list containing all terms of all s-halves.
\\ \\
\textcolor[HTML]{009B55}{\textbf{def}} \textcolor[HTML]{006EB8}{allS(sHalf,B):}\newline
Given a list of s-halves $sHalf$, and a list of critical pairs, compute all the s-halves. 
\\ \\
\textcolor[HTML]{009B55}{\textbf{def}} \textcolor[HTML]{006EB8}{macaulay(F):}\newline
Given a set of polynomials $F$, construct a sub-Macaulay matrix. In this function, we use a dictionary so that we get $O(1)$ lookup to get the index of each monomial in the matrix. For example if we have the polynomials $F = \{x^2 + x + y,y^2 + x\}$, the dictionary would look like this for monomial order $\DegRevLex$: $\text{colInd} = \{x^2:0,y^2:1, x:2,y:3\}$. We then construct an empty zero matrix of size $m \times n$, where $m$ is the number of polynomials and $n$ is the number of all unique monomials in $F$. If we take the first polynomial $x^2+x+y$, we get indices $\text{colInd[$x^2$]} = 0$, $\text{colInd[$x$]} = 2$, $\text{colInd[$y$]} = 3$, which are the indices where we will place the coefficients in our matrix. The result is then a sub-Macaulay matrix $\mathcal{M}_2(F)$.
\\ \\
\textcolor[HTML]{009B55}{\textbf{def}} \textcolor[HTML]{006EB8}{symPre(sHalf,G):}\newline
Symbolic preprocessing algorithm (Algorithm~\ref{symPre}). It takes a list of s-halves and the working Gröbner basis as input. It returns the list of s-halves, along with the new ones that have been computed. The sub-Macaulay matrix is not created in this function.   
\\ \\
\textcolor[HTML]{009B55}{\textbf{def}} \textcolor[HTML]{006EB8}{Red(sHalf,G,ver):}\newline
Reduction algorithm (Algorithm~\ref{Reduction}). It takes a list of s-halves, the working Gröbner basis and verbosity as input. It returns the list polynomials with \enquote{new} leading terms. The sub-Macaulay matrix is created in this function. If the verbosity ver is $1$ the printed output consists also of additional information. 
\\ \\
\textcolor[HTML]{009B55}{\textbf{def}} \textcolor[HTML]{006EB8}{F4(F,ver):}\newline
The main F4 algorithm (Algorithm~\ref{F4 alg}), which uses all of the functions described above. It returns a reduced Gröbner basis. If the verbosity ver is $1$ the printed output consists also of additional information. 

\subsection{A Toy Example}

We present a toy example of a Gr\"obner basis computation done with our Sage implementation of $\Ffour$. We consider the polynomial ring $R=\mathbb{F}_{101}[x,y,z]$ with the $\DegRevLex$ term order such that $x >_{\DRL} y >_{\DRL} z$.
We want to compute the reduced Gr\"obner basis of the ideal
$$(f_1 = x^3 + y^2 + xz - 1, f_2 = x^2+y^2+z-1, f_3 = y^2z + xz^2 - 1).$$
The input commands are the following:
\begin{verbatim}
K = GF(101)
variables = 3
names = 'x'
R = PolynomialRing(K, variables, names, order = 'degrevlex')
F = [R.0^3 + R.1^2 + R.0*R.2 - 1, R.0^2+R.1^2+R.2-1, R.1^2*R.2 + R.0*R.2^2 - 1]
F4(F,1) # high verbosity
\end{verbatim}

We describe what happens during the execution of the algorithm $\Ffour$.

Before entering the \textbf{while} loop and after the initial interreduction $\interred$, the polynomials of the working Gr\"obner basis are 
$$G^{(0)} = [f_1^{(0)} = xy^2 - y^2 - x + 1, f_2^{(0)} = x^2 + y^2 + z - 1, f_3^{(0)} = y^2z + xz^2 - 1],$$
i.e. the first polynomial $f_1$ in the list of generators has been replaced by the remainder of the multivariate division $\multiVarDiv$ between $f_1$ and $[f_2,f_3]$, made monic:
$$f_1^{(0)} = \frac{\multiVarDiv(f_1,[f_2,f_3])}{\LC({\multiVarDiv(f_1,[f_2,f_3]))}}, f_2^{(0)} = f_2, f_3^{(0)} = f_3.$$
Now, we enter in the \textbf{while} loop.

\subsubsection{First iteration of the \textbf{while} loop.}
We describe with great details what happens in the first iteration of the \textbf{while} loop.
We have $3$ possible pairs to consider, they are
$$P = [\Pair(f_1^{(0)},f_2^{(0)}), \Pair(f_1^{(0)},f_3^{(0)}),\Pair(f_2^{(0)},f_3^{(0)})].$$
The normal selection strategy selects the pairs of smallest degree and removes them from the list $P$. In this case, we have two such pairs. They are 
\begin{align*}
\Pair(f_1^{(0)},f_2^{(0)}) &= (x^2y^2,\\ & \ \ \ \ \ \shalf(f_1^{(0)},f_2^{(0)}) = x^2y^2 - xy^2 - x^2+ x,\\
&\ \ \ \ \ \shalf(f_2^{(0)},f_1^{(0)}) =  x^2y^2 + y^4 + y^2z - y^2),    
\end{align*}
and
\begin{align*}
\Pair(f_1^{(0)},f_3^{(0)}) &= (zy^2z,\\ & \ \ \ \ \ \shalf(f_1^{(0)},f_3^{(0)}) = xy^2z - y^2z - xz + z,\\
&\ \ \ \ \ \shalf(f_3^{(0)},f_1^{(0)}) =  xy^2z + x^2z^2 - x),    
\end{align*}

The Reduction $\Red$ will be done between
\begin{align*}
    L_1 &= [\shalf(f_1^{(0)},f_2^{(0)}) = x^2y^2 - xy^2 - x^2 + x,\\
    & \ \ \ \ \ \shalf(f_2^{(0)},f_1^{(0)}) = x^2y^2 + y^4 + y^2z - y^2,\\
    & \ \ \  \ \ \shalf(f_1^{(0)},f_3^{(0)}) = xy^2z - y^2z - xz + z,\\
    & \ \ \  \ \ \shalf(f_3^{(0)},f_1^{(0)}) =  xy^2z + x^2z^2 - x]
\end{align*} 
and $G^{(0)}$. In the Reduction step, we have the Symbolic Preprocessing $\symPre$, which is also done with inputs $L_1$ and $G^{(0)}$. The result is
\begin{align*}
    \symPre(L_1,G^{(0)}) &= [xy^2z - y^2z - xz + z, xy^2z + x^2z^2 - x \\
& \ \ \ \ \ \ x^2y^2 - xy^2 - x^2 + x, x^2y^2 + y^4 + y^2z - y^2 \\
& \ \ \ \ \ \ x^2 + y^2 + z - 1, y^2z + xz^2 - 1 \\
& \ \ \ \ \ \ xy^2 - y^2 - x + 1, x^2z^2 + y^2z^2 + z^3 - z^2],
\end{align*}
which can better be seen as a sub-matrix of the Macaulay matrix $\mathcal{M}_{\leq 4}$. In fact, if we put $M^{(1)} := \asmatrix(\symPre(L_1,G^{(0)}))$, we have that $M^{(1)}$ is equal to
$$\bordermatrix{
      & \textcolor{red}{x^2y^2} & y^4 & \textcolor{red}{xy^2z} & \textcolor{red}{x^2z^2} & y^2z^2 & \textcolor{red}{xy^2}  & \textcolor{red}{y^2z} & xz^2 & z^3  & \textcolor{red}{x^2}  & y^2  & xz & z^2  & x  & z  & 1  \cr
      & 0 & 0 & \textcolor{red}{1} & 1 & 0 & 0 &  0 & 0 & 0 & 0 & 0 &  0 & 0 & -1 & 0 & 0 \cr
      & 0 & 0 & \textcolor{red}{1} & 0 & 0 & 0 & -1 & 0 & 0 & 0 & 0 & -1 & 0 & 0 & 1 & 0 \cr
      & \textcolor{red}{1} & 1 & 0 & 0 & 0 &  0 & 1 & 0 & 0 & 0 & -1 & 0 & 0 & 0 & 0 & 0 \cr
      & \textcolor{red}{1} & 0 & 0 & 0 & 0 & -1 & 0 & 0 & 0 & -1 & 0 & 0 & 0 & 1 & 0 & 0 \cr
      & 0 & 0 & 0 & 0 & 0 &  0 & 0 & 0 & 0 & \textcolor{red}{1} & 1  & 0 & 0 & 0 & 1 & -1 \cr
      & 0 & 0 & 0 & 0 & 0 &  0 & \textcolor{red}{1} & 1 & 0 & 0 & 0  & 0 & 0 & 0 & 0 & -1 \cr
      & 0 & 0 & 0 & 0 & 0 & \textcolor{red}{1} & 0 & 0 & 0 & 0 & -1 & 0 & 0 & -1 & 0 & 1 \cr
      & 0 & 0 & 0 & \textcolor{red}{1} & 1 &  0 & 0 & 0 & 1 & 0 & 0  & 0 & -1& 0 & 0 & 0 \cr
}
,$$
where the rows are, respectively, from top to bottom
$$\shalf(f_1^{(0)},f_3^{(0)}),\shalf(f_3^{(0)},f_1^{(0)}),\shalf(f_1^{(0)},f_2^{(0)}),\shalf(f_2^{(0)},f_1^{(0)}),f_2^{(0)},f_3^{(0)},f_1^{(0)} \ \mbox{and} \ z^2 f_2^{(0)}.$$

In the Reduction step the RREF of $M^{(1)}$ is computed. We obtain that $\mbox{RREF}(M^{(1)})$ is
$$\bordermatrix{
  & \textcolor{red}{x^2y^2} & \textcolor{blue}{y^4} & \textcolor{red}{xy^2z} & \textcolor{red}{x^2z^2} & \textcolor{blue}{y^2z^2} & \textcolor{red}{xy^2}  & \textcolor{red}{y^2z} & xz^2 & z^3  & \textcolor{red}{x^2}  & y^2  & xz & z^2  & x  & z  & 1  \cr
  & \textcolor{red}{1} & 0 &  0 & 0 & 0 & 0 & 0 & 0 & 0 & 0 & 0 & 0 & 0 & 0 & 1 & 0 \cr
  & \textcolor{blue}{0} & \textcolor{blue}{1} & \textcolor{blue}{0} & \textcolor{blue}{0} & \textcolor{blue}{0} & \textcolor{blue}{0} & \textcolor{blue}{0} & \textcolor{blue}{-1} & \textcolor{blue}{0} & \textcolor{blue}{0} & \textcolor{blue}{-1} & \textcolor{blue}{0} & \textcolor{blue}{0} &
  \textcolor{blue}{0} & \textcolor{blue}{-1} & \textcolor{blue}{1} \cr
  & 0 & 0 & \textcolor{red}{1} & 0 & 0 & 0 & 0 & 1 & 0 & 0 & 0 & -1 & 0 & 0 & 1 & -1 \cr
  & 0 & 0 & 0 & \textcolor{red}{1} & 0 & 0 & 0 & -1 & 0 & 0 & 0 & 1 & 0 & -1 & -1 & 1 \cr
  & \textcolor{blue}{0} & \textcolor{blue}{0} & \textcolor{blue}{0} & \textcolor{blue}{0} & \textcolor{blue}{1} & \textcolor{blue}{0} & \textcolor{blue}{0} & \textcolor{blue}{1} & \textcolor{blue}{1} & \textcolor{blue}{0} & \textcolor{blue}{0} & \textcolor{blue}{-1} & \textcolor{blue}{-1} & \textcolor{blue}{1} & \textcolor{blue}{1} & \textcolor{blue}{-1} \cr
  & 0 & 0 & 0 & 0 & 0 & \textcolor{red}{1} & 0 & 0 & 0 & 0 & -1 & 0 & 0 & -1 & 0 & 1 \cr
  & 0 & 0 & 0 & 0 & 0 & 0 & \textcolor{red}{1} & 1 & 0 & 0 & 0 & 0 & 0 & 0 & 0 & -1 \cr
  & 0 & 0 & 0 & 0 & 0 & 0 & 0 & 0 & 0 & \textcolor{red}{1}
  & 1 & 0 & 0 & 0 & 1 & -1
}.$$
In \textcolor{blue}{blue} we have highlighted the polynomials
$$\textcolor{blue}{\widetilde{f_1^{(1)}}} = \textcolor{blue}{y^4 - xz^2 -y^2 - z + 1}, \textcolor{blue}{\widetilde{f_2^{(1)}}} = \textcolor{blue}{y^2z^2 + xz^2 + z^3 - xz - z^2 + x + z - 1},$$
which correspond to the new leading terms $\textcolor{blue}{y^4}$ and $\textcolor{blue}{y^2z^2}$ that we did not have before. Notice that the leading terms of the polynomials in $M^{(1)}$ are highlighted in \textcolor{red}{red}.
So, the result of the Reduction is precisely
$$\Tilde{F_1^+}= \Red(L_1,G^{(0)}) = [\widetilde{f_1^{(1)}} = y^4 - xz^2 - y^2 - z + 1, \widetilde{f_2^{(1)}} = y^2z^2 + xz^2 + z^3 - xz - z^2 + x + z - 1].$$

Now, before adding the results of the Reduction to the working Gr\"obner basis $G^{(0)}$, we interreduce them with $G^{(0)}$ and make them monic. This is done in order to build a working Gr\"obner basis which is as close as possible to the reduced one. 

Therefore, we compute $$f_{1,\bullet}^{(1)} := \frac{\multiVarDiv(\widetilde{f_1^{(1)}},G^{(0)})}{\LC(\multiVarDiv(\widetilde{f_1^{(1)}},G^{(0)}))},$$
which in this case yields again $\widetilde{f_1^{(1)}} (\not = 0)$, and we compute 
$$f_{2,\bullet}^{(1)} := \frac{\multiVarDiv(\widetilde{f_2^{(1)}},G^{(0)} \cup [f_{1,\bullet}^{(1)}])}{\LC(\multiVarDiv(\widetilde{f_2^{(1)}},G^{(0)} \cup [f_{1,\bullet}^{(1)}]))},$$
that yields $f_{2,\bullet}^{(1)} = xz^3 - xz^2 - z^3 + xz + z^2 - x - 2z + 1$.

Thus, the working Gr\"obner basis $G^{(1)}$ after round $1$ of the \textbf{while} loop is
\begin{align*}
G^{(1)} &= [f_1^{(0)},f_2^{(0)},f_3^{(0)},f_{1,\bullet}^{(1)},f_{2,\bullet}^{(1)}] \\
&= [-xy^2 + y^2 + x - 1, x^2 + y^2 + z - 1, y^2z + xz^2 - 1, \\
& \ \ \ \  \ \ y^4 - xz^2 - y^2 - z + 1, xz^3 - xz^2 - z^3 + xz + z^2 - x - 2z + 1] = \\
&= [f_1^{(1)},f_2^{(1)},f_3^{(1)},f_4^{(1)},f_5^{(1)}].
\end{align*}

Finally, we add $6$ new pairs to our list of pairs:
\begin{align*}
    P := P  \ \cup \ &[\Pair(f_1^{(0)},f_{1,\bullet}^{(1)}), \Pair(f_2^{(0)},f_{1,\bullet}^{(1)}),\Pair(f_3^{(0)},f_{1,\bullet}^{(1)}),\\
    & \ \Pair(f_1^{(0)},f_{2,\bullet}^{(1)}), \Pair(f_2^{(0)},f_{2,\bullet}^{(1)}),\Pair(f_3^{(0)},f_{2,\bullet}^{(1)})],
\end{align*}
which now consists of $7$ pairs.
This concludes the first iteration of the \textbf{while} loop. The step degree of this cycle is $4$, which is the highest degree of a polynomial in the sub-Macaulay matrix $M^{(1)}$.

\subsubsection{Other iterations of the \textbf{while} loop.}
There are $7$ more cycles of the \textbf{while} loop indexed by $i = 2,\dots,8$. In each iteration the same steps are performed, we outline them below.
\begin{itemize}
    \item Suppose $G^{(i-1)} = [f_1^{(i-1)},\dots,f_{m_{i-1}}^{(i-1)}]$, and that $P$ is the list of the pairs updated up to the execution of the $(i-1)$-th \textbf{while} loop.
    \item We select the pairs with the smallest degree according to the normal selection strategy, say
    $$P_i = \Sel(P) = [\Pair(f_s^{(i-1)},f_t^{(i-1)}) \ | \ (s,t) \in I_{i-1}],$$
    where $I_{i-1}$ is a set of bi-indexes. We remove the selected pairs from $P$.
    \item The Reduction $\Red$ will be done between
    $$L_i = [\shalf(f_s^{(i-1)},f_t^{(i-1)}), \shalf(f_t^{(i-1)},f_s^{(i-1)}) \ | \ (s,t) \in I_{i-1}]$$
    and $G^{(i-1)}$.
    \item In the Reduction step, we have the Symbolic Preprocessing $\symPre$, which is also done with inputs $L_i$ and $G^{(i-1)}$.
    \item We set $M^{(i)}=\asmatrix(\symPre(L_i,G^{(i-1)}))$ and compute
    $$\rows(\mbox{RREF}(M^{(i)}))$$
    and collect all the polynomials $\textcolor{blue}{\widetilde{f_1^{(i)}}},\dots,\textcolor{blue}{\widetilde{f_{n_i}^{(i)}}}$ whose leading terms did not appear as leading terms of polynomials in $\symPre(L_i,G^{(i-1)})$:
    $$\Tilde{F_i^+} = \Red(L_i,G^{(i-1)}) = [\widetilde{f_1^{(i)}},\dots,\widetilde{f_{n_i}^{(i)}}].$$
    \item Before adding these new polynomials to the working Gr\"obner basis, we perform a weak interreduction with $G^{(i-1)}$, i.e., for all $j=1,\dots,n_i$, we compute 
    $$f_{j,\bullet}^{(i)} := \frac{\multiVarDiv(\widetilde{f_j^{(i)}},G^{(i-1)} \cup ([f_{1,\bullet}^{(i)},\dots,f_{j-1,\bullet}^{(i)}] \smallsetminus [0]))}{\LC(\multiVarDiv(\widetilde{f_j^{(i)}},G^{(i-1)} \cup ([f_{1,\bullet}^{(i)},\dots,f_{j-1,\bullet}^{(i)}] \smallsetminus [0])))},$$ 
    and, if $f_{j,\bullet}^{(i)} \not= 0$, we add it to the working Gr\"obner basis. Thus, the new  working Gr\"obner basis is:
    \begin{align*}
    G^{(i)} &:= G^{(i-1)} \cup [f_{j,\bullet}^{(i)} \ | \ j = 1,\dots,n_i, \ f_{j,\bullet}^{(i)} \not= 0] \\
    &= [f_1^{(i)},\dots,f_{m_i}^{(i)}].    
    \end{align*}
    
    \item We now build all the new pairs of the form $\Pair(f,g)$ with $f \in G^{(i-1)}, g \in G^{(i)} \smallsetminus G^{(i-1)}$ and add it to the list $P$. This concludes the iteration.
    \item The step degree of this iteration $i$ of the \textbf{while} loop is the highest degree of an element in $\symPre(L_i,G^{(i-1)}).$
\end{itemize}
In Table~\ref{matrix:toyexample}, we report some relevant data collected during the execution of the \textbf{while} loop. These can be obtained by selecting $\mathrm{ver}=1$ when calling the function  \textcolor[HTML]{006EB8}{F4(F,ver)}.

\begin{table}[h]
   \centering
   \begin{tabular}{|c|c|c|c|c|c|}
   \hline
       iteration & step degree & $\#$ pairs & $\#$ sel.  pairs & $\#$ new pairs & size of $M^{(i)}$  \\
       \hline\hline
        1 & 4 & 3 & 2 & 6 & $8\times 16$ \\
        \hline
        2 & 5 & 7 & 4 & 5 & $19\times 25$ \\
        \hline
        3 & 3 & 8 & 2 & 6 & $5\times 7$\\
        \hline
        4 & 4 & 12 & 4 & 7 & $13\times 15$ \\
        \hline
        5 & 4 & 15 & 2 & 0 & $8\times 12$ \\
        \hline
        6 & 5 & 13 & 6 & 0 & $23\times 21$\\
        \hline
        7 & 6 & 7 & 5 & 0 & $29\times 32$ \\
        \hline
        8 & 7 & 2 & 2 & 0 & $16\times 26$\\
        \hline
   \end{tabular}
   \caption{For each iteration of the \textbf{while} loop, we report the step degree, the number of total pairs yet to be processed, the number of selected pairs, the number of new pairs obtained, and the size of the matrix $M^{(i)}$ which we 
 perform RREF over.}\label{matrix:toyexample}
   \label{tab:sizekeys}
\end{table}

At the end of the \textbf{while} loop, when we compute $G^{(8)}$, all pairs have been selected and no new pairs are added, so $G^{(8)}$ is a Gr\"obner basis. 
Finally, $G^{(8)}$ is interreduced using the $\interred$ function, and the reduced Gr\"obner basis we obtain is
$$[x^2 + z, y^2 - 1, xz^2 + z - 1, z^3 - xz + x].$$

%---------------------------------------------------------------------------------------------------------------------------------------------------

%BIBLIOGRAPHY
\newpage

\bibliography{biblio}
\bibliographystyle{siam}
\end{document}